%% file: main.tex
\documentclass[journal=mamobx,manuscript=article,layout=twocolumn]{achemso}

\input{header}

\modulolinenumbers[1]
\graphicspath{{figures}}
\SectionNumbersOn
\usepackage{xr}
\newboolean{boolnotes}
\newcommand{\Notes}[1]{\ifthenelse{\boolean{boolnotes}}{{\color{gray}#1}}{}}
\setboolean{boolnotes}{false}

\newcommand\crazeFig{0.55}
\newcommand\twoPlots{0.25}
\newcommand\threePlotsVer{0.24}

\author{Tobias Laschuetza}
\affiliation[Karlsruhe Institute of Technology]{Institute of Mechanics, Karlsruhe Institute of Technology, Kaiserstrasse 12, 76131 Karlsruhe, Germany}
\email{tobias.laschuetza@kit.edu}
\author{Ting Ge}
\affiliation[University of South Carolina]{Department of Chemistry and Biochemistry, University of South Carolina, Columbia, South Carolina 29208, United States}
\author{Thomas Seelig}
\affiliation[Karlsruhe Institute of Technology]{Institute of Mechanics, Karlsruhe Institute of Technology, Kaiserstrasse 12, 76131 Karlsruhe, Germany}
\author{Joerg Rottler}
\affiliation[University of British Columbia]{Quantum Matter Institute, University of British Columbia, Vancouver BC V6T 1Z4, Canada}
\alsoaffiliation{Departmentof Physics \& Astronomy,  University of British Columbia, Vancouver BC V6T 1Z1, Canada}

\title[Crazing]{Molecular simulations of crazes in glassy polymers under cyclic loading}

\abbreviations{MD}

\begin{document}
%\linenumbers

\input{sections/00_toc}

\begin{abstract}
We study with molecular dynamics simulations of a generic bead-spring model the cyclic crazing behaviour of glassy polymers. The aim is to elucidate the mechanical response of sole fibrillated craze matter as well as its interaction with bulk material. 
The macroscopic stress response exhibits a hysteresis, which is quasi stationary after the first cycle and largely independent of deformation rate and temperature. It results from a complex interplay between constraints imposed by the entanglement network, pore space and pore space closure.
Once the craze fibrils are oriented, stretching of the covalent backbone bonds leads to a rapid stress increase. In the initial stages of unloading, a loss in entanglement contact yields a quick stress relaxation in the backbone.
During unloading, the craze fibrils undergo a rigid body (i.e.\ stress-free) folding motion due to the surrounding pore space, so that the structural behaviour of craze fibrils during unloading is most accurately described as string-like. The reloading response depends significantly on the degree of pore space closure and the enforced intermolecular interaction during unloading. It ranges from a linear stress increase to a re-cavitation with a re-drawing response.
Compared to the bulk stiffness, the craze stiffness is two orders of magnitude lower and as a result, the macro response of coexisting craze and bulk matter is governed by the sole fibrillated craze matter.
\end{abstract}

\input{sections/01_intro}

\input{sections/02_model_methods}
\input{sections/03_results}
\input{sections/04_conclusions}
%\FloatBarrier
%\newpage

\input{sections/90_acknowledgement}

\input{sections/91_suppinfo}
%\FloatBarrier
%\newpage

%\bibliography{achemso-demo}
\bibliography{literature}

\end{document}

%% file: header.tex
\usepackage[T1]{fontenc} % Use modern font encodings
\usepackage{amsmath,amssymb,mathrsfs} % Helpdul math packages
\usepackage{xcolor}
\usepackage{lineno}
\usepackage{ifthen}
\usepackage{comment}

\usepackage{hyperref}
\usepackage[super]{nth}
\usepackage{stackengine}

\usepackage[section]{placeins} % allows the usage of \FloatBarrier
\usepackage[labelformat=simple]{subcaption} % enables \subcaptionbox

\renewcommand{\d}[1][]{\ensuremath{\,\mathrm{d}#1}}

\renewcommand{\phi}{\varphi}
\renewcommand{\epsilon}{\varepsilon}

\newcommand{\stretchMacro}{\ensuremath{\lambda_z}}
\newcommand{\stretchMacroMin}{\ensuremath{\lambda_{z,u}}}
\newcommand{\stretchMacroMax}{\ensuremath{\lambda_{z,max}}}
\newcommand{\extensionRatio}{\ensuremath{\lambda_{c}}}

\newcommand{\stress}{\ensuremath{\sigma_z}}
\newcommand{\stressPair}{\ensuremath{\sigma_{z,p}}}
\newcommand{\stressBond}{\ensuremath{\sigma_{z,b}}}

\newcommand{\orientation}{\ensuremath{S_z}}
\newcommand{\bondLength}{\ensuremath{N_i}}
\newcommand{\bondVector}{\ensuremath{\vec{R}(\bondLength, \stretchMacro)}}
\newcommand{\bondForce}{\ensuremath{\left< F_b \right>}}

\newcommand{\latParticleDisp}{\ensuremath{\left< u_{xy} \right>}}
\newcommand{\cycleTime}{\ensuremath{\tau_c}}

\newcommand{\LxIni}{\ensuremath{L_{x0}}}
\newcommand{\LyIni}{\ensuremath{L_{y0}}}
\newcommand{\LzIni}{\ensuremath{L_{z0}}}
\newcommand{\Lz}{\ensuremath{L_z}}
\newcommand{\Lc}[1][]{\ensuremath{L_{c #1}}}
\newcommand{\Li}[1][]{\ensuremath{L_{i #1}}}
\newcommand{\Lb}[1][]{\ensuremath{L_{b #1}}}

\newcommand{\eqcomma}{\ensuremath{\,\text{,}}}

%% file: sections/00_toc.tex
%%%%%%%%%%%%%%%%%%%%%%%%%%%%%%%%%%%%%%%%%%%%%%%%%%%%%%%%%%%%%%%%%%%%%%%%%
%%%%%%%%%%%%%%%%%%%%%%%%%%%%%%%%%%%%%%%%%%%%%%%%%%%%%%%%%%%%%%%%%%%%%%%%%

\begin{tocentry}
	\graphicspath{{figures/TOC}}
	\includegraphics[width=7.4cm]{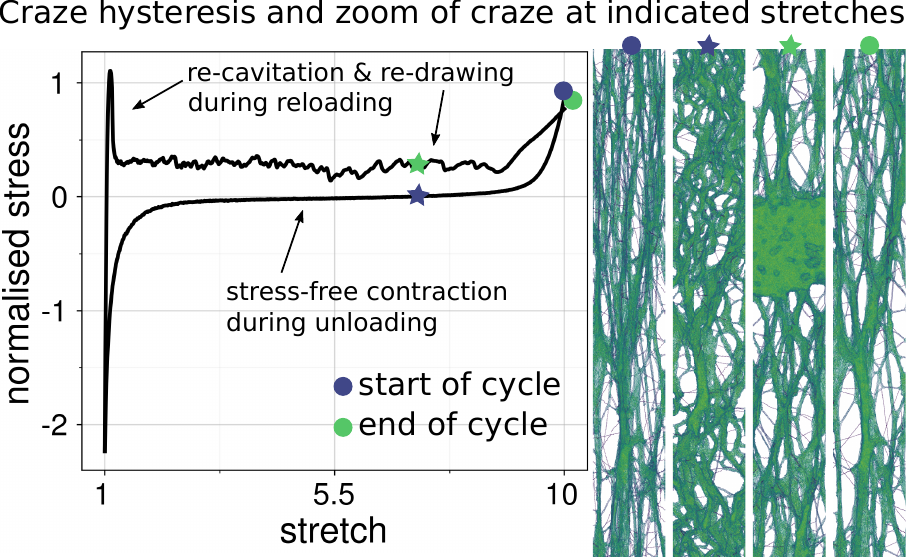}
%	\includegraphics[height=4.5cm]{TOC_2}

%	\includegraphics[height=\twoPlots\textheight]{stress_z_vs_stretch_z_21}

%	\begin{figure}[ht!]
%		\centering
%		\includegraphics[height=\twoPlots\textheight]{TOC}
%%		\caption{...}
%%		\label{fig_stress_split_5}
%	\end{figure}

%	Some journals require a graphical entry for the Table of Contents.
%	This should be laid out ``print ready'' so that the sizing of the
%	text is correct.
%	
%	Inside the \texttt{tocentry} environment, the font used is Helvetica
%	8\,pt, as required by \emph{Journal of the American Chemical
%		Society}.
%	
%	The surrounding frame is 9\,cm by 3.5\,cm, which is the maximum
%	permitted for  \emph{Journal of the American Chemical Society}
%	graphical table of content entries. The box will not resize if the
%	content is too big: instead it will overflow the edge of the box.
%	
%	This box and the associated title will always be printed on a separate page at the end of the document.
	
\end{tocentry}

%% file: sections/01_intro.tex
%%%%%%%%%%%%%%%%%%%%%%%%%%%%%%%%%%%%%%%%%%%%%%%%%%%%%%%%%%%%%%%%%%%%%%%%%
\section{Introduction}\label{sec_intro}
%%%%%%%%%%%%%%%%%%%%%%%%%%%%%%%%%%%%%%%%%%%%%%%%%%%%%%%%%%%%%%%%%%%%%%%%%
Crazing refers to the dilatant process of localized formation and growth of crack-like defects in glassy thermoplastic polymers. A craze consists of fibrillated matter with an interconnected void space. The several tens of nanometre thick fibrils can grow up to a few micrometers in length by drawing in surrounding bulk material from the so-called active zone \cite{Kramer.1990}. This process is of technical importance since, unlike cracks, craze fibrils enable a considerable load transfer between the craze surfaces and substantially enhance the fracture toughness. Therefore, much research has been devoted to understanding the governing mechanisms, cf.\ reviews in \cite{Kambour.1973,Kausch.1983,Kausch.1990,Haward.1997}, which includes theoretical studies on a continuum scale, e.g.\ \cite{Tijssens.2000a, Estevez.2000, Socrate.2001, Gearing.2004, Basu.2005, Seelig.2008, Helbig.2016}, as well as on a molecular scale, e.g.\ \cite{Baljon.2001, Rottler.2002.jamming, Rottler.2002, Rottler.2003, Basu.2010, toepperwein_2011,mahajan_2012, Venkatesan.2015, Ge.2014, Ge.2017, Wang.2022,dietz_2022,nan_2023}. 
However, the focus of those theoretical studies is limited to monotonic loading conditions. In contrast, extensive experimental research for cyclic (especially fatigue) loading exists, cf.\ reviews in  \cite{Doell.1990, Schirrer.1990, Takemori.1990}, providing insight into very interesting fracture processes for several glassy polymers and at several loading amplitudes: For instance, at low loading amplitudes, \citet{Skibo.1977} observed discontinuous crack growth, which was attributed to craze thickening resulting from a competition of fibril drawing and fibril creep deformation by \citet{Koenczoel.1990}. An increase in loading amplitude may give rise to the formation of shear bands and their interaction with crazes, leading to discontinuous epsilon-shaped fatigue cracks as studied by \citet{Takemori.1990}. 
Yet, the involved length scales pose difficulties to explore the driving mechanisms by solely relying on experiments and without theoretical analyses based on a physically motivated craze model. 
For this purpose, a continuum micromechanical model was recently developed \cite{Laschuetza.2024} focusing on the cyclic response of craze matter. The model describes the structural response of craze fibrils as string-like and accounts for viscoplastic fibril drawing and viscoelastic fibril deformation, where the latter is motivated by experimental observations \cite{Kambour.1969, Hoare.1972}. It is designed to be employed as a traction separation law in a mode I boundary value problem to investigate cyclic craze and crack growth. The continuum scale also allows to account for inelastic shear yielding in the surrounding bulk, which enables the analysis of its competition with crazing. 
However, the craze model suffers from two shortcomings arising from a general knowledge gap of the cyclic craze response: On the one hand, the structural behaviour of craze fibrils during unloading is uncertain. The correct structural behaviour is important since it impacts the pictures of the damage mechanism during cyclic (fatigue) loading. On the other hand, there is a lack in detailed mechanical knowledge regarding the response of craze deformation for a broad range of loading conditions.

In this paper, the response of a craze under cyclic loading is studied based on molecular dynamics simulations of a coarse-grained bead-spring model. In the past, this approach has yielded valuable insight into many molecular scale mechanisms of crazing in particular the ones concerning the entanglement network. General studies of craze nucleation \cite{Rottler.2003}, craze fibril drawing \cite{Baljon.2001,Rottler.2002.jamming,mahajan_2012,Venkatesan.2015}, and craze fracture \cite{Rottler.2002} in linear flexible polymers were extended to nanocomposites \cite{toepperwein_2011}, void nucleation \cite{Basu.2010}, semiflexible and stiff polymers \cite{dietz_2022,nan_2023}, and assessed the role of the entanglement network in setting the craze extension ratio \cite{Ge.2017,nan_2023}. 
Recent simulations have gone beyond the generic coarse-grained bead-spring model and used a scale-bridging approach to simulate craze formation in polystyrene \cite{Wang.2022}. 
Contrary to the energy landscape and stress level, the central role of the entanglement network in controlling the structural features of craze fibrils is independent of the degree of coarse graining.

The aim in this work is twofold: First, to the knowledge of the authors, molecular dynamics simulations have not yet been employed to study the cyclic response of fibrillated craze matter. The access to molecular scale details allows the simulations to establish the microscopic picture, which is often assumed in micromechanical models but is difficult to evaluate experimentally. Therefore, this study aims to elucidate the mechanical behaviour under cyclic loading and the underlying mechanisms leading to the macroscopic response. 
To this end, different loading conditions are analysed for sole fibrillated craze matter as well as for systems where bulk and craze matter coexist. The latter is of interest, since crazing is a transient process which gains particular relevance during cyclic loading where, for instance, fibril drawing takes place over multiple cycles. A further focus concerns the structural behaviour of craze fibrils during unloading and whether it can be characterised as string-like. 
Second, findings from this study aim to be transferred as a bottom-up approach to larger scale models to create, for instance, a molecular dynamics informed continuum model.

The presented paper is outlined as follows: In \autoref{sec_model_and_methods}, the computational methods featuring the model, the system setup and the crazing simulation are detailed. The results are presented in \autoref{sec_results} comprising the cyclic response of sole fibrillated craze matter (\autoref{sec_results_craze}) as well as the cyclic response of coexisting craze and bulk material for different craze-bulk compositions (\autoref{sec_results_craze_bulk}). Concluding, the key findings are summarised in \autoref{sec_conclusions}.

\Notes{
\bigskip
Extended motivation
\begin{itemize}
	\item Craze fibril properties difficult to access experimentally due to involved length scale
	\item Lack of understanding of craze (fibril) behaviour under cyclic loading
	\item Example of cyclic loading involves measurements of craze opening profile and then using integral methods or finite element simulations based on linear elastic bulk material to derive stress profile
	\item Not applicable to cyclic loading, if inelastic material behaviour exists
	\item Experimental observations such as retarded fatigue crack propagation or epsilon shaped fatigue cracks being a combination of crazing and shear yielding (inelastic isochoric deformation) not accessible
	\item Recently developed micro conti crazing model to investigate mode I fracture
	\item Fibril deformation viscoelastic and model assumption of string like fibrils
	\item Allows creep recovery while fibrils are loose hanging
	\item Verification of assumptions and obtain insight into actual structural behaviour of craze fibrils
\end{itemize}
}

\Notes{
Aim: craze response under cyclic loading, particularly focusing on:
\begin{itemize}
	\item Structural behaviour of craze fibrils during unloading: string-like vs.\ beam-like
	\item Driving mechanisms for cyclic response including hysteresis
	\item Bulk-craze interaction
\end{itemize}
}

%% file: sections/02_model_methods.tex
%%%%%%%%%%%%%%%%%%%%%%%%%%%%%%%%%%%%%%%%%%%%%%%%%%%%%%%%%%%%%%%%%%%%%%%%%
\section{Model and methods}\label{sec_model_and_methods}
%%%%%%%%%%%%%%%%%%%%%%%%%%%%%%%%%%%%%%%%%%%%%%%%%%%%%%%%%%%%%%%%%%%%%%%%%
The generic bead-spring model \cite{Kremer.1990} is employed to study the cyclic craze response.
%, since it has proven effective in capturing the process of craze formation, growth and breakdown \cite{Rottler.2003} and is qualitatively in good agreement with results obtained from more detailed united atom simulations on crazing \cite{Wang.2022}. 
The model setup is consistent with previous studies on crazing, e.g.\ \cite{Baljon.2001, Rottler.2002.jamming, Rottler.2002, Rottler.2003, Ge.2014, Ge.2017}, and is briefly summarised: Each polymer is modelled as a linear chain of $N$ spherical monomers with a mass $m$. Non-bonded monomers separated by the distance $r$ interact via the truncated and shifted 6-12 Lennard-Jones (LJ) potential
\begin{footnotesize}
\begin{equation}
	U_{\textrm{LJ}}(r) = 4 u_0 [\left( a/r \right)^{12} - \left( a/r \right)^{6} - \left( a/r_c \right)^{12} + \left( a/r_c \right)^{6} ] 
\end{equation}
\end{footnotesize}
for $r \leq r_c$ and
%
%\begin{equation}
%	U_{\textrm{LJ}}(r) = 4 u_0 \left[ \left( \frac{a}{r} \right)^{12} - \left( \frac{a}{r} \right)^{6} - \left( \frac{a}{r_c} \right)^{12} + \left( \frac{a}{r_c} \right)^{6} \right] \qquad \textrm{for} \quad r \leq r_c
%\end{equation}
with a cutoff radius $r_c=1.5a$. The results are expressed in terms of Lennard-Jones units, featuring the characteristic length $a$, energy scale $u_0$ and a characteristic time $\tau = a \sqrt{m/u_0}$. The bonded monomers interact via an attractive finitely extensible nonlinear elastic (FENE) potential with a purely repulsive second Lennard Jones term ($r_c=2^{1/6} a$)
\begin{footnotesize}
\begin{equation}
	U_{\textrm{FENE}}(r) = -\frac{1}{2} k R_0^2 \ln [ 1 - \left( r/R_0 \right)^2 ] +  4 u_0 [ \left( a/r \right)^{12} - \left( a/r \right)^{6}]
\end{equation}
\end{footnotesize}
%\begin{equation}
%	U_{\textrm{FENE}}(r) = -\frac{1}{2} k R_0^2 \ln\left[ 1 - \left( \frac{r}{R_0} \right)^2 \right] +  4 u_0 \left[ \left( \frac{a}{r} \right)^{12} - \left( \frac{a}{r} \right)^{6} \right]
%\end{equation}
with the parameters $R_0=1.5a$ and $k=30 u_0/a^2$, allowing entanglements to form  \cite{Kremer.1990}. Chain scission is not considered, since the focus lies on cyclic loading resulting in stress states lower than necessary for chain scission to occur \cite{Rottler.2003}.

The system comprises $M=1800$ chains where each chain consists of $N=500$ beads with an average entanglement length of $N_e \approx 80$ beads. Simulations are performed with the LAMMPS molecular dynamics code \cite{LAMMPS.1995,LAMMPS.2022}, where periodic boundary conditions are employed along all three directions. 

The melt state was constructed by generating random-walk coils with a subsequent equilibration with $r_c=1.12a$ at temperature $T=1 u_0/k_b$. To facilitate equilibration, a double bridging algorithm \cite{Auhl.2003} is utilized. The temperature was controlled via a Nosé-Hoover thermostat with a damping rate $1 \tau^{-1}$. While retaining the volume, the equilibrated system is then quenched with $r_c=1.5 a$ to $T=0.49 u_0/k_b$, leading to zero pressure. Further quenching to the final target temperature $T=0.1 \, u_0/k_b$ takes place at zero pressure by using a Nosé-Hoover barostat with a damping rate $0.1 \tau^{-1}$. The quenched system has initial box dimensions $\LxIni=94.3a$ and $\LyIni=55.8a$ in lateral direction and $\LzIni=167.3a$ in axial direction. All directions sufficiently exceed the end-to-end distance of the chains, which reduces finite size effects \cite{Rottler.2003, Hoy.2005}. The larger box length along the z-direction minimises further the effects of the finite box size on the coexistence of the craze fibrils and un-crazed regions. $\LzIni$ being larger than the active zone size allows better separation of the craze fibrils from the active zone at the same axial stretch $\stretchMacro=\Lz/\LzIni$, which is measured with respect to the isotropic glass. The box sizes in the periodic x- and y-directions are reduced compared to $\LzIni$, so the simulations with the same number of degrees of freedom can focus on the more interesting z-direction. 

Following previous studies, e.g.\ \cite{Rottler.2002, Rottler.2003,Ge.2017,Wang.2022}, crazing is induced by subjecting the system to an affine uniaxial deformation along the $z$-axis, i.e.\ lateral stretches $\lambda_x = \lambda_y = 1$, while $\stretchMacro$ is prescribed via a constant material strain rate $\dot{\lambda}_z = \dot{u}/\LzIni= 2.6 \cdot 10^{-4} \tau$. To simulate cyclic loading conditions, a bi-linear loading/unloading program is imposed by simply switching the velocity direction upon reaching the maximum stretch $\stretchMacroMax$, leading to the loading period $T_p$. Throughout the loading program, a Nosé-Hoover thermostat with a damping rate $1 \tau^{-1}$ is applied to the lateral velocities to maintain a constant temperature.

To study the cyclic response of sole craze material in \autoref{sec_results_craze}, the isotropic glass is deformed to $\stretchMacroMax=10$. This ensures complete conversion of bulk material into fibrillated craze matter, which takes place when the deformation reaches approximately the extension ratio $\extensionRatio$ (i.e.\ $\stretchMacro \approx \extensionRatio =\rho_b/\rho_c$), describing the ratio of bulk $\rho_b$ and craze density $\rho_c$. It also avoids the deformation state where chain breakage occurs (cf.\ \cite{Rottler.2003}).
In \autoref{sec_results_craze_bulk}, the bulk-craze interaction for different bulk-craze compositions is studied by varying $\stretchMacroMax$. To analyse a broad response range, the unloading magnitude $\stretchMacroMin$, i.e.\ the stretch to which the system is unloading, is varied in both studies. In all considered scenarios, the cycle count and the cycle time $\tau_c=\tau/T_p$ commences upon reaching $\stretchMacroMax$ for the first time after loading from the isotropic glass.

\Notes{
\begin{itemize}
	\item Coarse-grained bead spring model of $M=1800$ linear chains where each polymer is a chain of $N=500$ spherical monomers with monomer mass $m$
	\item System is quenched to a temperature $T=0.1 \, u_0/k_b$
	\item Lennard-Jones (LJ) units: molecular diameter $a$, binding energy $u_0$, characteristic time $\tau = a \sqrt{m/u_0}$
	\item Bi-linear loading/unloading programme with uniaxial deformation, i.e.\ $\lambda_x = \lambda_y = 1$, and constant material strain rate $\dot{\lambda}_z = \dfrac{\dot{u}}{l_z(t=0)}= 2.6 \cdot 10^{-4} \tau$
	\item Initial monotonic deformation to stretch $\stretchMacroMax$ 
	\item In first part of this paper $\stretchMacroMax=10$ ensuring complete conversion of bulk material into fibrilated craze matter. Subsequent cyclic loading investigates three different unloading magnitudes of $\stretchMacroMin=1$, $\stretchMacroMin=2$ and $\stretchMacroMin=5$
	\item In second part $\stretchMacroMax=5$ allowing to study the craze-bulk interaction
	\item Cycle count w.r.t.\ crazed configuration at $\stretchMacroMax$, which is defined by the cycle time $\tau_c=t/T_p$ where $T_p/\tau$ is the loading period
\end{itemize}
}

%% file: sections/03_results.tex
%%%%%%%%%%%%%%%%%%%%%%%%%%%%%%%%%%%%%%%%%%%%%%%%%%%%%%%%%%%%%%%%%%%%%%%%%
\section{Results}\label{sec_results}
\graphicspath{{figures/results}}
%%%%%%%%%%%%%%%%%%%%%%%%%%%%%%%%%%%%%%%%%%%%%%%%%%%%%%%%%%%%%%%%%%%%%%%%%
\subsection{Cyclic response of fibrillated craze matter} \label{sec_results_craze}
\begin{figure}[ht!]
	\centering
	\includegraphics[height=0.145\textheight]{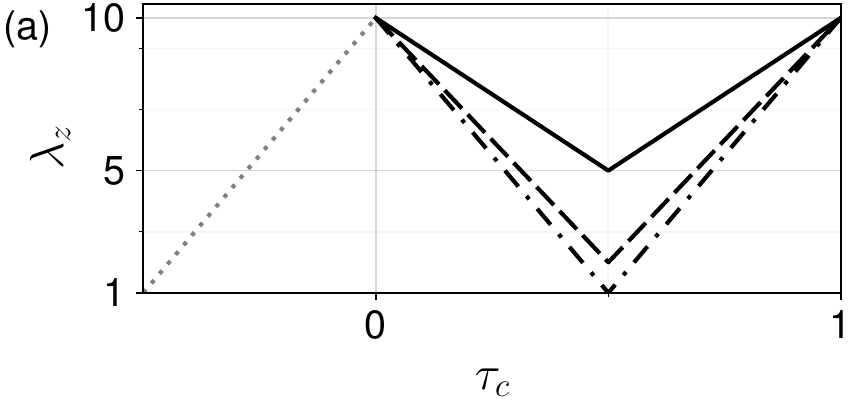}
	{\phantomsubcaption\label{fig_stress_stretch_load}}
	\\ \vspace*{0.5em}
	\includegraphics[height=\twoPlots\textheight]{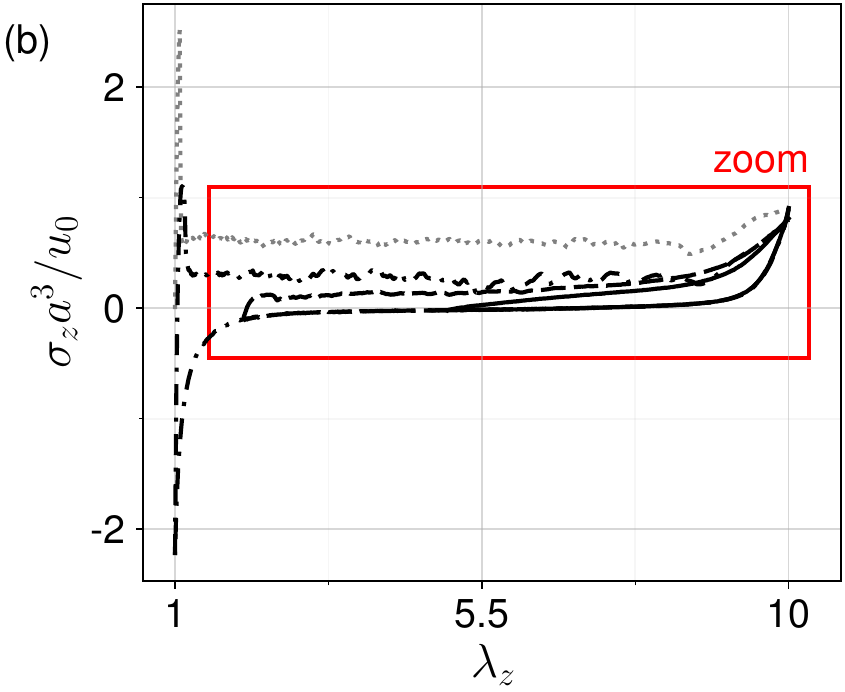}
	{\phantomsubcaption\label{fig_stress_stretch_a}}
	\\ \vspace*{0.5em}
	\includegraphics[height=\twoPlots\textheight]{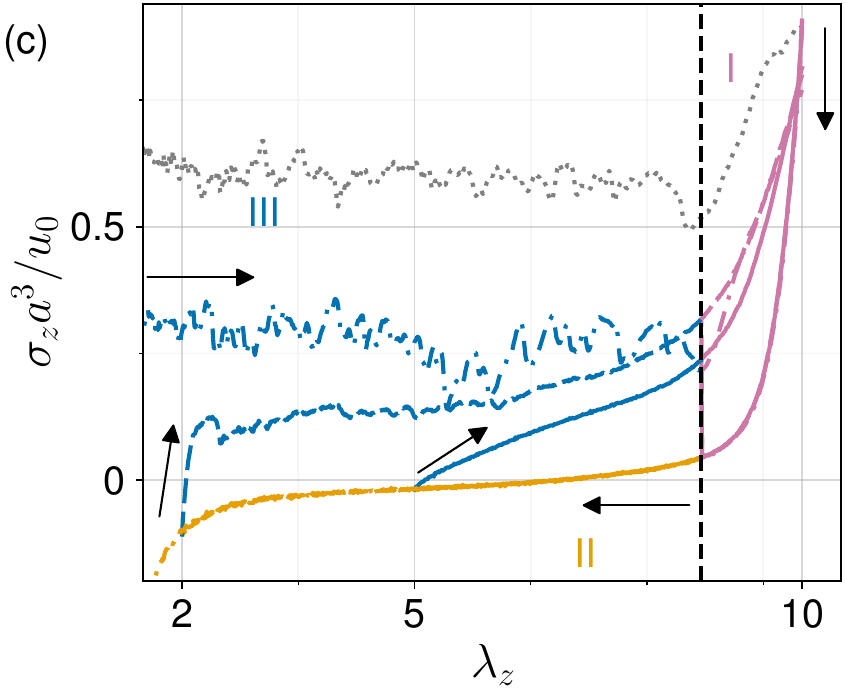}
	{\phantomsubcaption\label{fig_stress_stretch_zoom}}
	\caption{Mechanical response during initial craze formation (grey dotted line) and subsequent cyclic loading (black lines) to various unloading stretches of $\stretchMacroMin=1$, $\stretchMacroMin=2$ and $\stretchMacroMin=5$, represented by black dash-dotted, dashed, and solid lines, respectively. \subref*{fig_stress_stretch_load} Schematic loading program, \subref*{fig_stress_stretch_a} axial stress $\stress$ as function of axial stretch $\stretchMacro$ with red box indicating zoom shown in \subref*{fig_stress_stretch_zoom}. The arrows and colour coding in panel \subref*{fig_stress_stretch_zoom} show loading direction and different craze stages, respectively. The latter are described in the text.}
	\label{fig_stress_stretch}
\end{figure}

We first aim to elucidate the mechanical response of sole craze matter by focusing on the macroscopic behaviour throughout the loading cycles and thereafter analysing the underlying mechanisms. \autoref{fig_stress_stretch} presents the macroscopic response of fibrillated craze matter under cyclic loading for three simulations to $\stretchMacroMin=1$ (dash-dotted line), $\stretchMacroMin=2$ (dashed line) and $\stretchMacroMin=5$ (solid line). The deformation controlled cyclic loading program is shown in \autoref{fig_stress_stretch_load}, and the corresponding macroscopic axial stress $\stress$ as function of the macro stretch $\stretchMacro$ is depicted in \autoref{fig_stress_stretch_a}. The grey dotted line represents the initial response of the isotropic glass during craze formation and growth, featuring an elastic stress increase, the subsequent stress drop during cavitation, as well as a constant stress plateau as craze fibrils are drawn from the bulk. This monotonic crazing process has already been thoroughly studied, e.g.\ \cite{Baljon.2001, Rottler.2002, Rottler.2003, Ge.2017}.
Here we are interested in the mechanical response of the craze (fibrils) under cyclic loading (black lines). To provide a better visualisation, \autoref{fig_stress_stretch_zoom} shows a zoom of the stress-stretch response (red box in \autoref{fig_stress_stretch_a}), and the arrows in \autoref{fig_stress_stretch_zoom} show the loading direction. The colour coding and the roman numerals indicate different craze stages.

%\begin{figure}[ht!]
%	\centering
%	\subcaptionbox{
%		\label{fig_stress_stretch_load}
%	}{
%		\includegraphics[height=0.145\textheight]{stretch_z_vs_cycle}
%	}\\
%	\subcaptionbox{
%		\label{fig_stress_stretch_a}
%	}{
%		\includegraphics[height=\twoPlots\textheight]{stress_z_vs_stretch_z_21_22_25_with_zoom}
%	}\\
%	\subcaptionbox{
%		\label{fig_stress_stretch_zoom}
%	}{
%		\includegraphics[height=\twoPlots\textheight]{stress_z_vs_stretch_z_21_22_25_zoom_color_coding}
%	}
%	\caption{\jr{It would look nicer to have labels (a) (b) (c) inside top left corners. In this and all other figures in the draft, the font for the axis labels is too small for my taste.} Mechanical response during initial craze formation (grey dotted line) and subsequent cyclic loading (black lines) to various unloading stretches of $\stretchMacroMin=1$, $\stretchMacroMin=2$ and $\stretchMacroMin=5$, represented by black dash-dotted, dashed, and solid lines, respectively. \subref*{fig_stress_stretch_load} Schematic loading program, \subref*{fig_stress_stretch_a} axial stress $\stress$ as function of axial stretch $\stretchMacro$ with red box indicating zoom shown in \subref*{fig_stress_stretch_zoom}. The arrows and colour coding in panel \subref*{fig_stress_stretch_zoom} show loading direction and different craze stages, respectively. The latter are described in the text.}
%	\label{fig_stress_stretch}
%\end{figure}

\Notes{
Key observations:
\begin{itemize}
	\item Cyclic loading leads to hysteresis, which is neither a thermally activated nor rate-dependent process (cf.\ Figure S1)
	\item 3 stages can be distinguished in the mechanical response as indicated in \autoref{fig_stress_stretch_zoom}, namely:
	\begin{itemize}
		\item Stage 1: stretching of highly oriented craze fibrils, which is denoted here as the first stage as response is essentially independent of the previous loading history (cf.\ \autoref{fig_stress_stretch_zoom}).
		\item Stage 2: quasi stress-free contraction of the craze with a transition to compression around $\stretchMacro=2.5$
		\item Stage 3: a reloading response showing, depending on the unloading magnitude $\stretchMacroMin$, either a re-yielding ( $\stretchMacroMin \leq 2$) or a linear ($\stretchMacroMin \geq 5$) behaviour. 
	\end{itemize}
    \item Subsequent analysis focuses on the \nth{1} cycle if not indicated differently
\end{itemize}
}

Before discussing the different stages of the craze deformation cycle, several key features are briefly summarised. During unloading from $\stretchMacroMax=10$ (indicated by the vertical arrow in \autoref{fig_stress_stretch_zoom}, the craze first exhibits a rapid stress decline which transitions into a long, quasi-stress free contraction of the craze (orange line). When the stretch is reduced below $\stretchMacroMin < 2.5$, the quasi stress-free contraction becomes compressive, and the magnitude of $\stress$ significantly rises as $\stretchMacro$ approaches $\stretchMacro=1$ (dash-dotted line), i.e.\ the macroscopically undeformed state of the initial isotropic glass. In that particular case, $\stress$ increases rapidly during reloading and the craze exhibits what appears to be re-cavitation and re-yielding, while both stress magnitudes remain below the cavitation and drawing stresses during the initial craze creation (grey dotted line). By contrast, for the case of $\stretchMacroMin=2$ (dashed line) and $\stretchMacroMin=5$ (solid line), the craze exhibits a plateau stress and a linear increase, respectively, during reloading (cf.\ blue dashed and solid line in \autoref{fig_stress_stretch_zoom}). Around $\stretchMacro \approx \extensionRatio$, which is indicated by the vertical line in \autoref{fig_stress_stretch_zoom}, the stress rapidly grows and the response becomes nearly independent of $\stretchMacroMin$. Moreover, it is notable that all load program variations lead to a hysteresis, which appears to be neither a thermally activated nor a rate-dependent process and relatively stationary after the first cycle (cf.\ Figure S1 and Figure S2 in the supporting information). 
%A hysteresis was also observed in experimental studies focusing on craze deformation under cyclic loading \cite{Kambour.1969, Hoare.1972}. 

%Yet, the underlying mechanisms appear to be different. This is based on two observations: First, the significant difference in time scales and second, successive cyclic loading as shown in Figure S2, leads to a relatively stationary hysteresis. In contrast, the pure uniform deformation of a polycarbonate craze without drawing in new craze material shows the tendency to an elastic response upon successive loading \cite{Kambour.1969}. \jr{We need to discuss this paragraph via zoom, I am not sure I understand and agree with the claims.}

To analyse the driving mechanisms leading to the hysteresis, it is beneficial to distinguish between different craze stages during the deformation cycle. As aforementioned, the stages are represented by the colour-coding and the roman numerals in \autoref{fig_stress_stretch_zoom} and can be summarised as follows: 
\begin{itemize}
	\item Stage I (magenta): Stretching and unloading of highly oriented craze fibrils. This is labelled here as first stage, since the response is essentially independent of the previous loading history.
	\item Stage II (orange): Quasi stress-free contraction of the craze with a transition to compression around $\stretchMacro=2.5$.
	\item Stage III (blue): A reloading response which depends on $\stretchMacroMin$ and exhibits either a re-yielding ($\stretchMacroMin \leq 2$) or a linear ($\stretchMacroMin \geq 5$) behaviour.
\end{itemize}
To further enhance the understanding, animations (created with OVITO \cite{OVITO.2010}) of the three simulations are provided in the supporting information.

The further analysis focuses on the \nth{1} cycle as defined in \autoref{fig_stress_stretch_load}, if not indicated otherwise. 

%\FloatBarrier
\subsubsection{Mechanics during stage I}
\begin{figure}[ht!]
	\centering
	\includegraphics[height=\twoPlots\textheight]{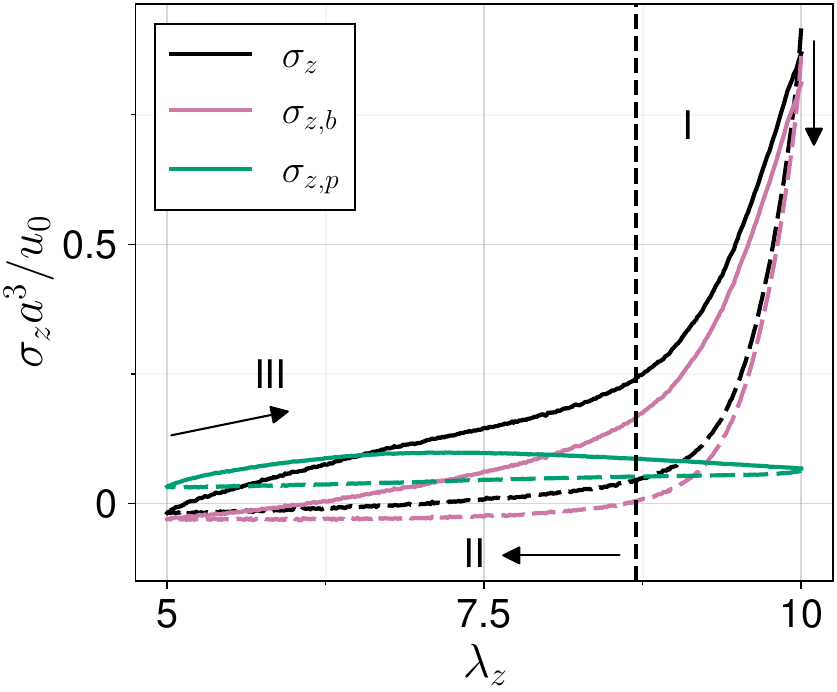}
	\caption{Axial stress $\stress$ decomposition into intermolecular pair $\stressPair$ and intramolecular bond $\stressBond$ components during unloading (dashed lines) and reloading (solid lines) for $\stretchMacroMin=5$. Arrows show load direction.}
	\label{fig_stress_split_5}
\end{figure}

The mechanics in stage I are discussed first while focusing on the mechanisms leading to the rapid stress increase during reloading, and the difference between loading and unloading that results in the hysteresis. The former is investigated through a decomposition of $\stress$ into its intramolecular $\stressBond$ and intermolecular $\stressPair$ contributions, as shown in \autoref{fig_stress_split_5} for $\stretchMacroMin=5$. Since the kinetic contribution is small in the glass, it is omitted here. Stress decompositions for the cases $\stretchMacroMin=2$ and $\stretchMacroMin=1$ are shown in Figure S3 and exhibit qualitatively similar trends. Focusing only on stage I (i.e.\ right hand-side of vertical dashed line), it can be concluded that the rapid stress change is governed by $\stressBond$ and hence, by the deformation of the polymer backbone. 

\begin{figure}[ht!]
	\centering
	\includegraphics[height=\twoPlots\textheight]{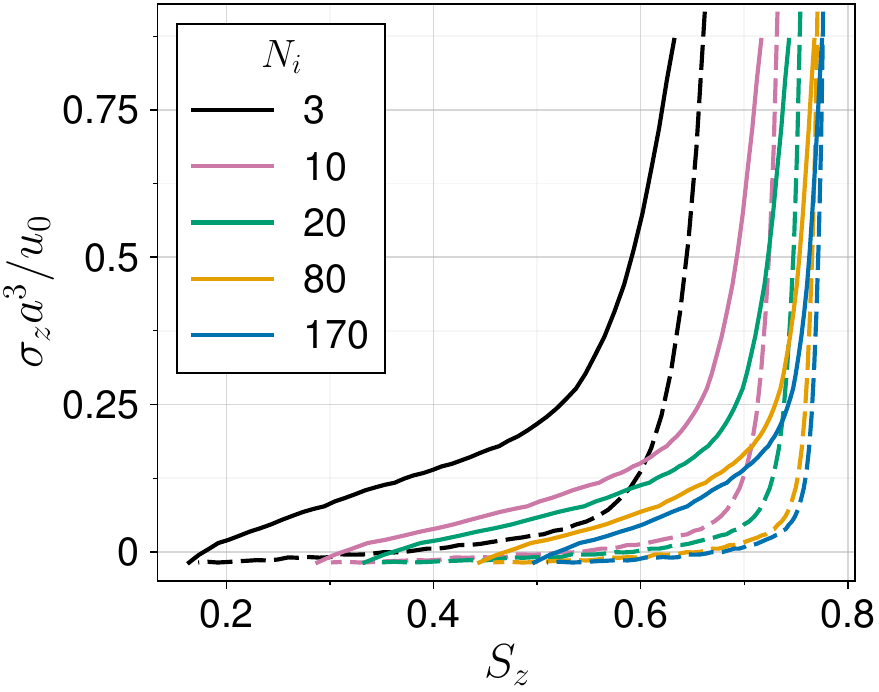}
	\caption{Axial stress $\stress$ vs.\ bond orientation $\orientation$ for several length scales encompassing $\bondLength$ monomers. Data is shown during unloading (dashed lines) and reloading (solid lines) for $\stretchMacroMin=5$.}
	\label{fig_P2}
\end{figure}

Since \autoref{fig_stress_split_5} suggests that chain (re)orientation is important, we further analyze the changes in chain configuration by computing orientation vectors $\bondVector$ between $N_i$ beads along the chains for a given stretch $\lambda_z$. The average orientation of these vectors is conveniently described by
\begin{equation}
	\orientation = \left< P_2(\cos \alpha_z(\bondLength, \stretchMacro)) \right> \eqcomma
\end{equation}
where $P_2(x)=(3x^2-1)/2$ is the second Legendre polynomial, $\alpha_z$ the angle formed by $\bondVector$ with the deformation axis and $\left< \right>$ describes the ensemble average.
\autoref{fig_P2} shows a parametric plot of the stress as function of $\orientation $ for five values of $N_i$, where $N_i=80$ coincides with the entanglement length $N_e$.   

The first notable observation is that $\orientation$ is not an adequate order parameter at any length scale $N_i$ to characterise the hysteresis, since the hysteresis loops do not collapse. Second, given that $\orientation(\bondLength=80) \approx \orientation(\bondLength=170)$, the orientation seems to be enforced by the entanglement network. Third, before unloading from $\stretchMacroMax$, the chains are highly oriented, which is essentially retained during the initial stages of unloading (dashed lines) in which the stress rapidly decays. Lastly,  the rapid stress rise during reloading (solid lines) in regime I is accompanied by a minor change in orientation. 

\begin{figure}[ht!]
	\centering
	\includegraphics[height=\twoPlots\textheight]{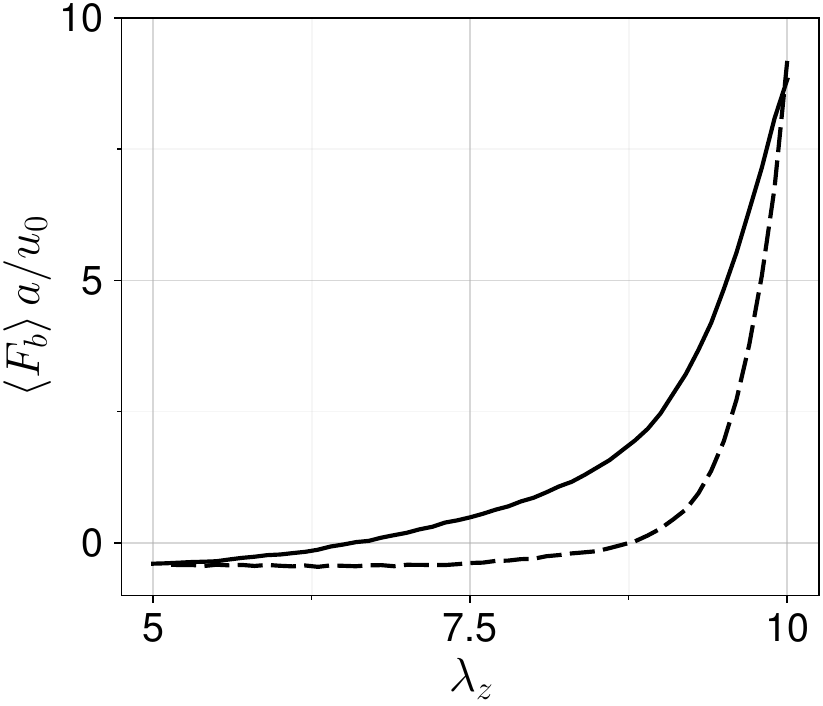}
	\caption{Average backbone bond force $\bondForce$ vs.\ stretch $\stretchMacro$ during unloading (dashed lines) and reloading (solid lines) for $\stretchMacroMin=5$.}
	\label{fig_avg_bond_force}
\end{figure}

The stretching and relaxing of the backbone bonds is further corroborated by \autoref{fig_avg_bond_force}, which shows the average resultant tensile bond force $\bondForce = \left< \d U_{\textrm{FENE}}(r)/\d r \right>$. It closely follows the bond stress $\stressBond$ derived from the virial stress, and paints the picture that constraints such as the entanglement network leads to a rise in $\stress$ once chains become highly oriented. Furthermore, upon unloading, the loss in entanglement contacts yields a swift stress relaxation in the backbone and thus causes the difference in loading and unloading.

In conclusion, stage I is governed by the stretching and relaxing of the highly oriented bonds due the entanglement network. As the bonds form craze fibrils, it equivalently can be stated that this stage is governed by the deformation of the highly oriented craze fibrils. Given the necessary pre-orientation in order for the deformation to take place, this stage is essentially independent of the prior unloading history (i.e.\ of $\stretchMacroMin$). 

\Notes{
\begin{itemize}
	\item During stage I, chains are highly oriented (cf.\ \autoref{fig_P2}) and response is essentially independent of the previous loading history (cf.\ \autoref{fig_stress_stretch_zoom})
 	\item Main contribution of stress due to bond interaction (cf.\ \autoref{fig_stress_split_5})
	\item Entanglement contact leads at further deformation to a rapid increase in bond stress (cf.\ \autoref{fig_avg_bond_force})
	\item Upon unloading, loss of entanglement contact leads to quick stress $\stress$ relaxation
\end{itemize}
}

%\FloatBarrier
\subsubsection{Mechanics during stage II}
Stage II focuses on the mechanisms leading to the quasi stress-free contraction of the craze during unloading as well as the transition to compression for $\stretchMacro \leq 2.5$. The data presented so far highlights two things in this regime: Firstly, the largest reduction in chain orientation occurs during this stage (cf.\ \autoref{fig_P2}) and secondly, $\stressBond \approx \bondForce \approx 0$ (cf.\ \autoref{fig_stress_split_5} and \autoref{fig_avg_bond_force}). The latter suggests that the structural behaviour of craze fibrils is string-like rather than beam-like as the craze (pore space \& fibrils) solely exhibits a minor stiffness during contraction. This also implies a negligible bending stiffness of the craze fibrils, which validates a recent assumption on the structural behaviour of craze fibrils used in a continuum model \cite{Laschuetza.2024}.

\begin{figure}[ht!]
	\centering
	\includegraphics[height=\twoPlots\textheight]{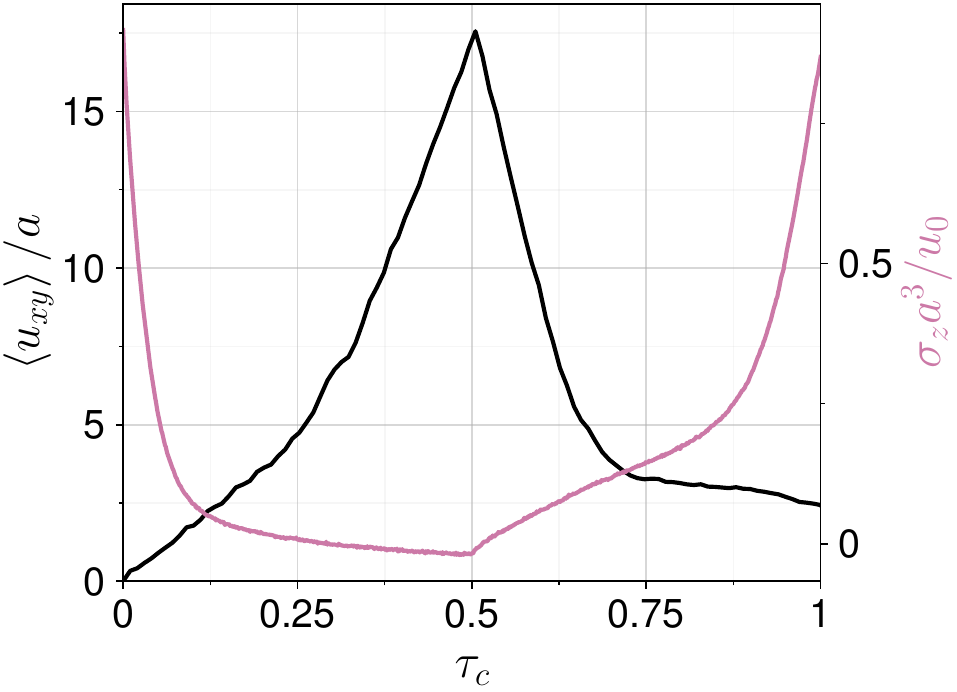}
	\caption{Lateral particle displacement $\latParticleDisp$ (left y-axis) and $\stress$ (right y-axis) vs.\ cycle time $\cycleTime$ for $\stretchMacroMin=5$.}
	\label{fig_u_xy}
\end{figure}

Besides the chain level observables, a further helpful quantity to characterise the response is the particle motion orthogonal (i.e.\ in $xy$-plane) to the deformation axis ($z$-axis). Since $\lambda_x=\lambda_y=1$, any such motion is nonaffine. The average lateral particle displacement $\latParticleDisp$ is given by 
\begin{equation}
	\latParticleDisp = \left< \left\lvert \vec{r}_{xy}(\cycleTime) - \vec{r}_{xy}(\cycleTime=0) \right\rvert \right> \eqcomma
\end{equation}
where $\vec{r}_{xy}$ is the position vector in the $xy$-plane and $\latParticleDisp$ is the displacement with respect to $\stretchMacroMax$ at the beginning of the cycle at $\cycleTime=0$. \autoref{fig_u_xy} displays the evolution of $\latParticleDisp$ (black line) as well as $\stress$ (magenta line) throughout the loading cycle. During unloading ($0 < \cycleTime \leq 0.5$), $\latParticleDisp$ increases sharply. That is, the lateral particle movement increases while the bond orientation decreases, which can also be seen in the supplementary animation as a "folding mechanism" of the craze fibrils. 
The pore space plays a key role as it enables the motion of fibril folding to occur essentially as a rigid body motion, i.e.\ stress-free. It is therefore concluded, that as long as pore space exists, the structural behaviour of craze fibrils is thus most accurately described as string-like.

The compressive stress increase observed for $\stretchMacro \leq 2.5$ (cf.\ \autoref{fig_stress_stretch_a}) arises from an increase in $\stressPair$ (cf.\ stress decomposition of $\stretchMacroMin=1$ and $\stretchMacroMin=2$ in Figure S3). Although the magnitude is quite large and most likely strongly influenced by the high deformation speed in MD simulations, this behaviour is not unexpected due to the macroscopically observed dilatant deformation during craze formation and growth, leading to compressive stress ($\stress < 0 $) during unloading while the macroscopic deformation is still tensile ($\stretchMacro>1$). Compression at the crack tip, i.e.\ where the craze is the largest, was also computed using the experimentally measured craze contour as input in a finite element simulation \cite{Bevan.1986, Koenczoel.1990}.

\Notes{
\begin{itemize}
	\item Key feature in stage II is the stress-free contraction of the craze during unloading highly implying a string-like response of the craze, which is in contrast to an assumed beam-like response (\textbf{ADD SOURCE})
	\item Animation shows a strong motion (rotation and displacement) of the craze fibrils 
	\item Data also shows a reduction of bond orientation on all length scales (cf.\ \autoref{fig_P2}) and a elevated lateral particle movement (cf.\ \autoref{fig_stage_3})
	\item The combination of a string-like craze fibril behaviour and existing pore space enables stress-free craze contraction
	\item Once the unloading magnitude reaches $\stretchMacroMin \leq 2$, pore space closure is enforced and the LJ interaction yields a high macroscopic compressive force (cf.\ stress split in Figure S3)
\end{itemize}
}

%\FloatBarrier
\subsubsection{Mechanics during stage III}
\begin{figure*}[!ht]
	\hfil%
	\begin{minipage}[b]{0.4\textwidth}
		\flushright
		\stackunder{
			\includegraphics[height=\crazeFig\textheight]{../craze/21_stretch_10_1_lam_5_annotated_2}
			{\phantomsubcaption\label{fig_craze_lengths_1}}
		}{
			$\stretchMacroMin=1$
		}
		\hfil
		\stackunder{
			\includegraphics[height=\crazeFig\textheight]{../craze/25_stretch_10_2_lam_5}
			{\phantomsubcaption\label{fig_craze_lengths_2}}
		}{
			$\stretchMacroMin=2$
		}
		\hfil
		\stackunder{
			\includegraphics[height=\crazeFig\textheight]{../craze/22_stretch_10_5_lam_5}
			{\phantomsubcaption\label{fig_craze_lengths_5}}
		}{
			$\stretchMacroMin=5$
		}
	\end{minipage}
	\hfil%
	\begin{minipage}[b]{0.4\textwidth}
		\flushleft
		\includegraphics[width=\threePlotsVer\textheight]{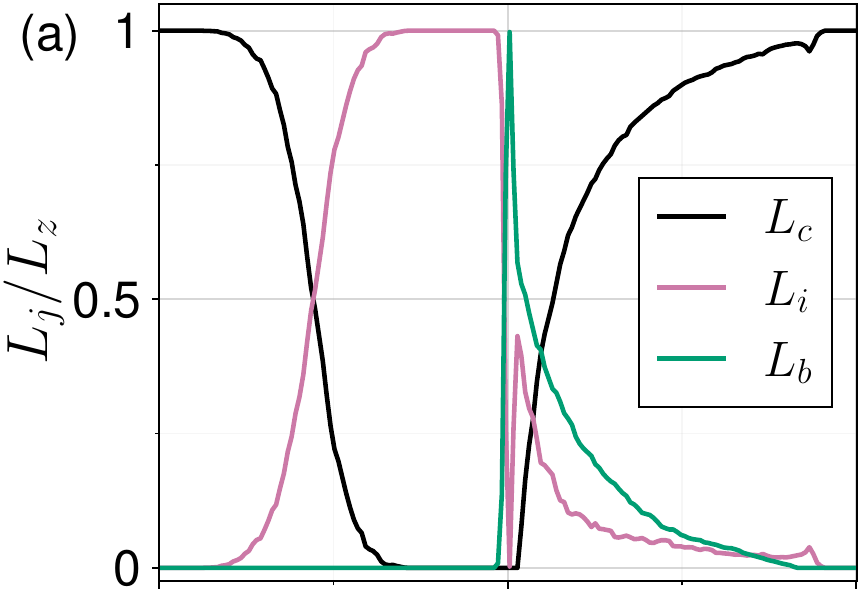}
		\\ \vspace*{0.5em}
		\includegraphics[width=\threePlotsVer\textheight]{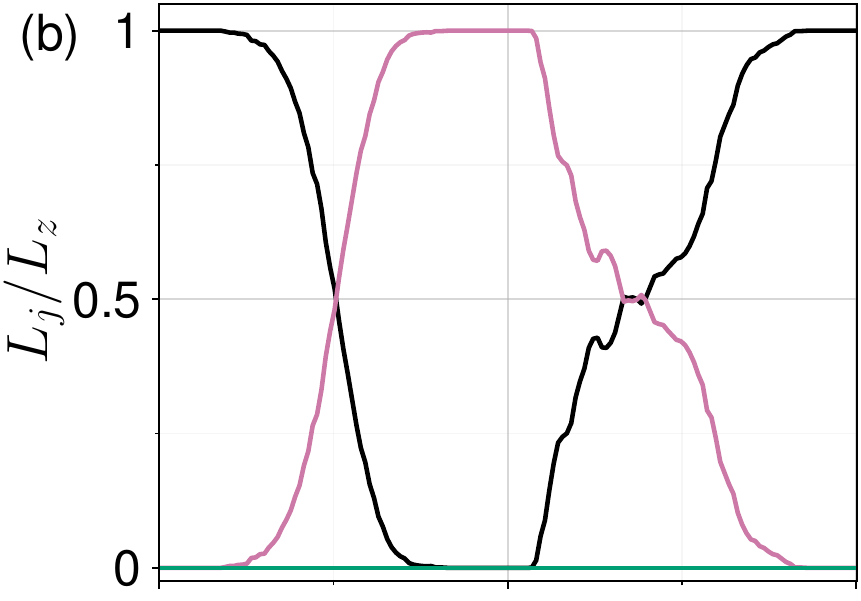}
		\\ \vspace*{0.5em}
		\includegraphics[width=\threePlotsVer\textheight]{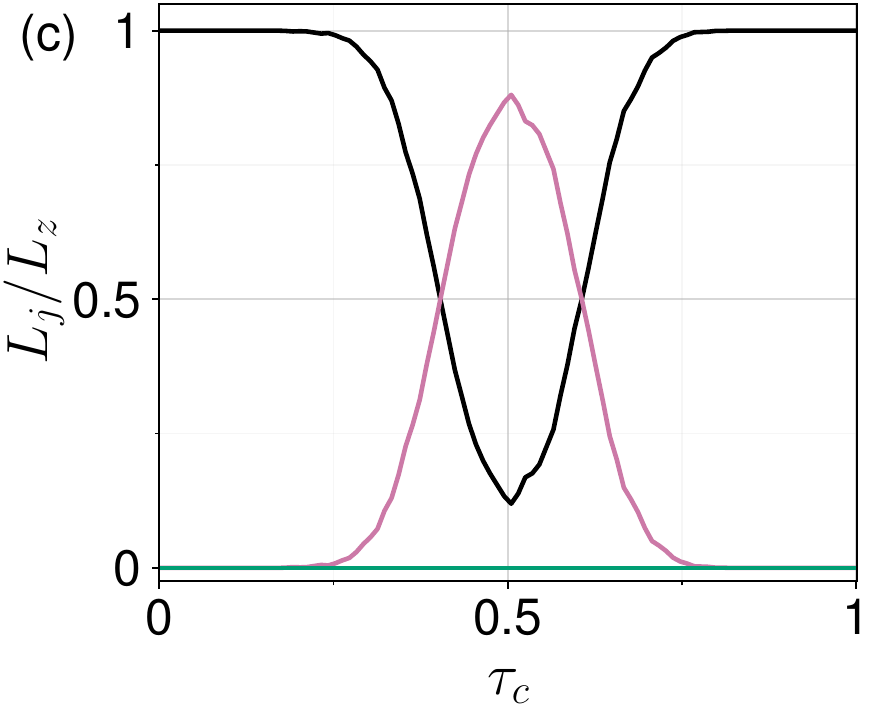}
	\end{minipage}\hfil
	\caption{Snapshots of crazes during reloading at $\stretchMacro=5$ and corresponding evolution of craze length components throughout cycle for \subref*{fig_craze_lengths_1} $\stretchMacroMin=1$, \subref*{fig_craze_lengths_2} $\stretchMacroMin=2$ and \subref*{fig_craze_lengths_5} $\stretchMacroMin=5$.}
	\label{fig_craze_lengths}
\end{figure*}

Stage III features the reloading response, which significantly depends on $\stretchMacroMin$. The supplementary animations show that in the case of $\stretchMacroMin=5$, the craze morphology is largely retained throughout the cycles. By contrast, $\stretchMacroMin=2$ yields a section-wise unfolding of the craze and $\stretchMacroMin=1$ even a re-drawing. The previous data implies that pore space closure and the exerted compression highly influence the reloading behaviour. To further study the underlying mechanism that contributes especially to the difference between $\stretchMacroMin=2$ and $\stretchMacroMin=1$, the simulation box length $\Lz$ is decomposed into regions associated with the bulk material
\begin{equation}
	\Lb = \int_{\{z: \rho(\cycleTime,z) \geq \rho_b\}} \d z \eqcomma
\end{equation}
the craze
\begin{equation}
	\Lc = \int_{\{z: \rho(\cycleTime,z) \leq \rho_c\}} \d z
\end{equation}
and an intermediate length
\begin{equation}
	\Li = \int_{\{z: \rho_c < \rho(\cycleTime,z) < \rho_b\}} \d z \eqcomma
\end{equation}
where $\rho(\cycleTime,z)$ is the density of a slab $\d z$ at $\cycleTime$ and $\rho_b$ and $\rho_c$ are the bulk and craze density, respectively. The evolution of those craze length components is displayed in \autoref{fig_craze_lengths}, including additionally snapshots of the craze (created with OVITO \cite{OVITO.2010}) for the three simulations compressed to different $\stretchMacroMin$-values, and then restrained to the same $\stretchMacro=5$. To further facilitate the understanding of the approach, the area enclosed by the coloured dashed lines in snapshot \autoref{fig_craze_lengths_1} defines the three contributions associated with $\Lb$, $\Lc$ and $\Li$.

In all three simulations, the simulation box consists solely of crazed material ($\Lc = \Lz$) at $\cycleTime=0$,. As the deformation is retracted and the craze unloaded, $\Li$ increases monotonically to $\Lz$. In the case of $\stretchMacroMin = 1$, the high compressive stress eventually leads to a very rapid increase from $\Lb=0$ to $\Lb=1$ (cf.\ green line around $\cycleTime \approx 0.5$ in \autoref{fig_craze_lengths_1}). Upon reloading, $\Li$ quickly rises and subsequently $\Lb$ as well as $\Li$ continuously decrease. This is associated with the re-drawing of fibrils as can be seen in the animation. The small rise of $\Li$ can be understood as new "active zone" (magenta area shown in the \autoref{fig_craze_lengths_1} snapshot), from which fibrils are drawn. In this context, it is important to realize that the decrease in $\Li$ results merely from the normalisation, but is rather a constant active zone length (see Figure S5 in a non-normalised plot for $\Li$). We also note that it would be misleading to characterise this behaviour as healing of the craze, as neither the temperature nor the simulation time allows for any substantial chain reptation. The response arises rather from the breakage of intermolecular interactions, which were created during $\stress < 0$ (cf.\ Figure S3(a)).

This behavior is in strong contrast to $\stretchMacroMin \geq 2$, where $\Lb$ remains zero. While the key feature of $\stretchMacroMin=5$ is the retention of the craze morphology seen by smooth transitions between $\Lc$ and $\Li$ in \autoref{fig_craze_lengths_5}, $\stretchMacroMin=2$ is more interesting. With the system reloading from $\Li=1$ at $\cycleTime=0.5$ (\autoref{fig_craze_lengths_2}), re-cavitation does not occur and the breaking of LJ interactions (cf.\ Figure S3(b)) results in a relatively constant plateau stress (cf.\ \autoref{fig_stress_stretch_zoom}). The animation shows an "unfolding" behaviour of the craze, which can bee seen by the evolution around $ 0.5< \cycleTime < 0.8$ in \autoref{fig_craze_lengths_2}. This becomes even more evident if the normalisation is omitted and $\Li$ and $L_c$ are plotted separately as shown in Figure S4.

%\tl{Potentially comment on initial stiffness for $\stretchMacroMin=5$ due to chain orientation.}

\Notes{
\begin{itemize}
	\item Stage III is characterised by a reloading response, which depends whether unloading caused pore space closure
	\item If reloading commences while pore space still exists (e.g.\ for $\stretchMacroMin \geq 5$), then the linear stress increase in this stage is attributed to the increase in bond orientation while bond orientation has returned to the same level as at the beginning of the prior loading cycle 
	\item Craze exhibits pronounced craze morphology memory throughout cyclic loading if no pore space closure occurs (cf.\ animations)
	\item On the other hand, pore space closure leads to an increase in LJ interaction (Figure S3) 	
	\item Based on the density profile of the craze, the length of the craze can be distinguished between a bulk $\Lb = \int_{\{z: \rho(z) \geq \rho_b\}} dz$, craze $\Lc = \int_{\{z: \rho(z) \leq \rho_c\}} dz$ and intermediate contribution $\Li = \int_{\{z: \rho_c < \rho(z) < \rho_b\}} dz$, which are displayed in \autoref{fig_craze_lengths}	
	\item In case of $\stretchMacroMin = 1$, the system is compressed leading to $\Lb=1$ at the beginning of the cycle. Subsequent deformation yields a continuous reduction in bulk density (\autoref{fig_craze_lengths_1}) which is associated with the re-drawing response occurring during reloading (cf.\ animation). This is the only investigated case where $\Lb>0$ after the first cycle.
	\item In contrast, $\stretchMacroMin=2$ gives rise to a section wise unfolding of the craze (cf.\ animation). With the initial system reloading from $\Li=1$ (\autoref{fig_craze_lengths_2}), re-cavitation does not occur and the break of LJ interactions result in a relatively constant plateau stress (cf.\ \autoref{fig_stress_stretch_zoom}).  
\end{itemize}
}

%\FloatBarrier
\subsection{Bulk-craze interaction} \label{sec_results_craze_bulk}
\begin{figure}[ht!]
	\centering
	\includegraphics[height=\twoPlots\textheight]{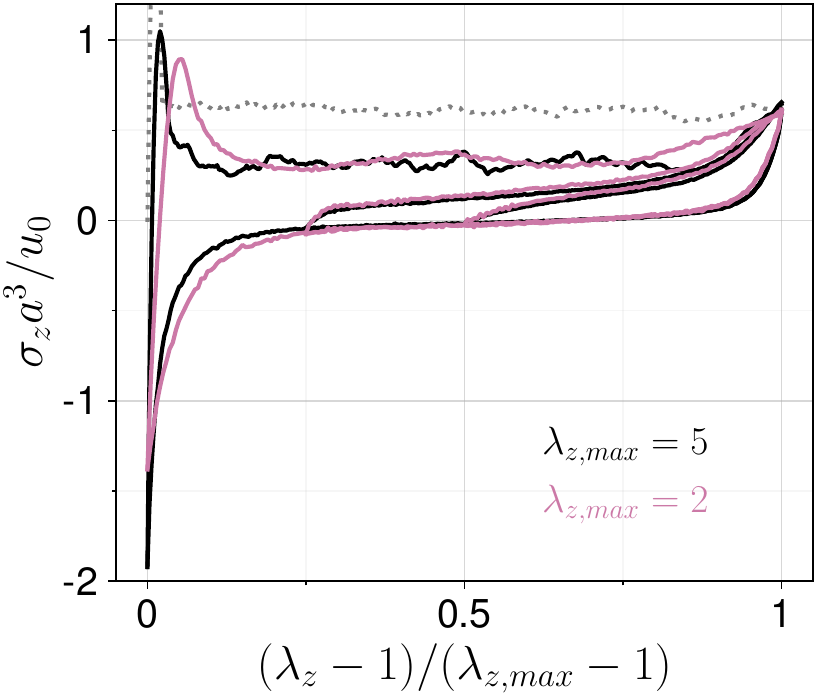}
	\caption{Stress-strain curves for two bulk-craze configurations obtained by $\stretchMacroMax=5$ (black lines) and $\stretchMacroMax=2$ (magenta lines) as well as the initial craze formation (grey dotted line). Each configuration features three unloading levels $\stretchMacroMin$.}
	\label{fig_bulk_craze_int}
\end{figure}

The study so far has focused on the cyclic deformation  of fully crazed matter, which has been drawn from the isotropic glass. Yet, fibril drawing is a transient process and hence it is also interesting to consider the cyclic response while craze and bulk material coexist. To investigate the role of the craze length, two configurations are created by deforming the initial glass to $\stretchMacroMax=5$ as well as $\stretchMacroMax=2$, leading to craze/bulk length ratios at $\stretchMacroMax$ of $\Lc/\Lb=9.8$ and $\Lc/\Lb=1.3$, respectively. The protocol then follows the previous protocol by commencing the cyclic loading routine to three different $\stretchMacroMin$-values, leading to the bulk-craze interaction for the nine simulations shown in \autoref{fig_bulk_craze_int}. The deformation has been rescaled by $\epsilon_z = (\stretchMacro-1)/(\stretchMacroMax-1)$, resulting in a collapse of the curves for a given unloading $\epsilon_{z,u} = (\stretchMacroMin-1)/(\stretchMacroMax-1)$. 

The bulk-craze interaction leads to a response very similar to the pure craze response in \autoref{fig_stress_stretch} even for "short" crazes such as $\Lc/\Lb \approx 1$. This allows for two important conclusions: Firstly, the macro deformation is governed by the craze and the bulk plays a minor role in cyclic deformation. This is attributed to the much higher bulk stiffness $E_b$ compared to the craze stiffness $E_c$. To further elaborate that, the bulk/craze stiffness can be crudely estimated by their respective secant stiffness with respect to $\stretchMacro$ while neglecting 3D effects. This yields $E_b/E_c \approx 100$, where $E_b$ is evaluated at $\stretchMacro=(1, 1.5)$ and $E_c$ for reloading in regime I at $\stretchMacro=(9.5, 10)$ in \autoref{fig_stress_stretch_zoom}. It is important to acknowledge that this is the craze stiffness and not the craze fibril stiffness. Secondly, the stress free contraction occurring also for shorter craze fibrils corroborates a negligible bending stiffness and hence the string-like structural response of craze fibrils. 

We conclude this section by discussing an estimate of the craze length $\xi_0$ at $\stretchMacroMax$. This estimate is limited to the case without pore space closure (i.e.\ $\epsilon_z \geq 0.25$), which is then motivated by the scaling in \autoref{fig_bulk_craze_int}. 
%Taking note of the layered bulk-craze structure, assuming small elastic deformations and using continuum micromechanical arguments, such as mass balance during fibril drawing and kinematic coupling between the layers, (for details also including finite strains see \cite{Laschuetza.2024}) yields 
%\begin{equation}
%	\xi_0 = h_0 (\stretchMacroMax - 1)\dfrac{\extensionRatio}{\extensionRatio-1} \eqcomma
%\end{equation}\jr{Are you quoting a result derived elsewhere? If not, some more explanation should be given, I at least do not understand this equation.}
%where $h_0$ is the primordial thickness, which is here taken as $h_0 = \Lz(\tau=0)$. Denoting the estimated bulk layer by this approach as $\chi_0$ leads to $\xi_0/\chi_0=10$ and $\xi_0/\chi_0=1.3$ for $\stretchMacroMax=5$ and $\stretchMacroMax=2$, respectively, which is in good agreement with $\Lc/\Lb$. As a final remark, it is noteworthy that no continuous drawing is observed on the investigated time scales, which would be an indication for viscous drawing or viscous deformation effects of the fibril.
While drawing is not exhausted, the craze essentially exhibits a layered bulk-craze structure at $\stretchMacroMax$ in which the deformed craze layer is denoted by $\xi$ and the deformed bulk layer by $\chi$. They differ from $\Lc$ and $\Lb$ as $\xi$ and $\chi$ are calculated in the following and not given by the MD simulations. Using further continuum micromechanical arguments (for details including finite strains see \cite{Laschuetza.2024}) yields the kinematic coupling between the macro stretch $\stretchMacro$ and the layers $\xi$ and $\chi$ as
\begin{equation} \label{eq_kinematic_coupling}
	\stretchMacro = \dfrac{\chi + \xi}{h_0} \eqcomma
\end{equation} 
where $h_0 = L_{z0}$ is the primordial thickness. Furthermore, the unloaded bulk $\chi_0$ and craze length $\xi_0$ are related by conservation of mass during the drawing process, given by
\begin{equation} \label{eq_cons_mass}
	\chi_0 = h_0 - \xi_0 /\extensionRatio \eqcomma
\end{equation}
where $\extensionRatio$ is the extension ratio with respect to the unloaded configuration.
To combine \eqref{eq_kinematic_coupling} and \eqref{eq_cons_mass} without proposing constitutive deformation models, we assume small (elastic) deformations of the bulk and craze, i.e.\ $\chi_0 \approx \chi$ and $\xi_0 \approx \xi$, allowing to derive an estimation for the craze fibril length
\begin{equation} \label{eq_fibril_length_est}
	\xi_0 = h_0 (\stretchMacroMax - 1)\dfrac{\extensionRatio}{\extensionRatio-1} 
\end{equation}
at $\stretchMacroMax$.
Applying \eqref{eq_fibril_length_est} to $\stretchMacroMax=5$ and $\stretchMacroMax=2$ leads to $\xi_0/\chi_0=10$ and $\xi_0/\chi_0=1.3$, respectively, which is in good agreement with $\Lc/\Lb$ values given by the MD simulations. 

As a final remark, it is noteworthy that no continuous drawing is observed on the investigated time scales, which would be an indication for viscous drawing or viscous deformation effects of the fibril.

\Notes{
\begin{itemize}
	\item Study of bulk-craze interaction
	\item Crazing process not completed in \nth{1} cycle
	\item Response similar to sole craze response, i.e.\ deformation dominated by craze
	\item Stress free contraction also for shorter craze fibrils corroborating a negligible bending stiffness
	\item No continuous drawing on investigated time scale, potentially relaxation sim for more insight
\end{itemize}
}

%% file: sections/04_conclusions.tex
%%%%%%%%%%%%%%%%%%%%%%%%%%%%%%%%%%%%%%%%%%%%%%%%%%%%%%%%%%%%%%%%%%%%%%%%%
\section{Conclusions}\label{sec_conclusions}
\graphicspath{{figures/results}}
%%%%%%%%%%%%%%%%%%%%%%%%%%%%%%%%%%%%%%%%%%%%%%%%%%%%%%%%%%%%%%%%%%%%%%%%%
Molecular dynamics simulations were used to study the mechanical response of sole fibrillated craze matter and the bulk-craze interaction in glassy polymers under cyclic loading. The maximum loading amplitude $\stretchMacroMax$ and unloading magnitude $\stretchMacroMin$ were varied to simulate different bulk-craze compositions and different degrees of pore space closure, respectively. A key finding of this study was that in all investigated cases, i.e.\ the sole craze and the craze-bulk system, the macro responses exhibited a hysteresis, which was found to be quasi stationary after the first cycle and largely independent of deformation rate and temperature. 
%This is somewhat in contrast to the experimentally observed craze fibril behaviour \cite{Kambour.1969}, which showed a tendency to a purely elastic response upon reloading. 

Another salient finding was that the hysteresis cannot be simply described by the chain orientation \orientation, but rather resulted from a complex interplay between constraints imposed by the entanglement network, pore space and pore space closure. Three distinct craze stages were identified to study the driving mechanism in detail: 
The first stage focused on the highly oriented bonds and craze fibrils. Further deformation led to an accelerated increase in stress $\stress$ (cf.\ \autoref{fig_stress_stretch}), which resulted from the stretching of the covalent backbone bonds. Upon unloading the oriented craze, the bond force $\bondForce$ quickly relaxed due to the loss in entanglement contact, yielding a rapid drop in $\stress$. 
The second stage characterised the stress-free contraction during further unloading, in which the animations showed that the craze fibrils undergo a folding motion. The folding motion was characterised on the chain and particle level observables by a decrease in chain orientation and elevated lateral particle movement, respectively. The surrounding pore space is the essential trait for the folding motion to take place as a rigid body motion (i.e.\ stress-free). 
The third stage described the reloading response, which exhibited a strong dependency on $\stretchMacroMin$. It was shown that a key feature is the degree of pore space closure and the necessary intermolecular interaction to enforce it. Complete pore space closure ($\stretchMacroMin=1$) was accompanied by a high level of intermolecular stress, which led to a re-cavitation and re-drawing during reloading. If the pore space was maintained throughout the cyclic (e.g.\ $\stretchMacroMin=5$), the reloading response was bi-linear.

The study on the craze-bulk interaction revealed that even "short" craze fibrils, where $\Lb \approx \Lc$, exhibited the three characteristic stages of sole fibrillated craze matter described above. That is, once a craze initiates and reaches a certain length, the macroscopic response is governed by the craze. This was attributed to the significantly lower craze stiffness (with respect to $\stretchMacro$) of $E_b/E_c \approx 100$. It also allows to use continuum micromechanical considerations to estimate the craze-bulk ratio at peak loading.

A further important finding, there has not been an indication of a macroscopic bending stiffness during unloading, even in the case of "short" craze fibrils. The folding motion occurs on the craze fibril level which comprises locally stiff craze fibril segments (cf.\ animations). Therefore, it is likely that the results qualitatively hold for semiflexible bead-spring models, i.e.\ with a bending potential. 
While such models primarily yield a higher entanglement density, the crazing process is still dilatant by creating pore space and hence, the mechanisms leading to the stress-free contraction of the craze are still present. Concluding, the macroscopic structural response of craze fibrils during unloading is most accurately described as string-like, despite locally stiff craze fibril segments. This finding is important, since a craze fibril bending stiffness would have different implications on the damage mechanisms during cyclic (fatigue) loading.

The most interesting contribution for future work includes carefully conducted experiments on fibrillated craze matter investigating the response under different loading conditions. 
%A valuable contribution for future work include a constitutive craze fibril model which allows to incorporate the response on a continuum scale. 

\Notes{
\begin{itemize}
	\item By varying the unloading magnitude $\stretchMacroMin$, different degrees of pore space closure are simulated 
	\item In all cases the craze exhibits a hysteresis, which has been found to to be largely independent of deformation speed and temperature
	\item A key finding is that the hysteresis results from an complex interplay between entanglement contact, pore space and pore space closure and that it cannot be described by the chain orientation \orientation, since it does not collapse when stress is plotted versus \orientation
	\item The progressive stress increase occurs once chains are oriented in the deformation direction and entanglement contact then leads to a rapid deformation of the bonds upon further stretching
	\item Upon unloading, the rapid stress drop is associated with a loss in entanglement contact leading to a rapid relaxation of the stress along the bonds
	\item During unloading, the craze fibrils undergo a folding motion which is accompanied by a decrease in chain orientation and elevated lateral particle movement and occurs essentially as a rigid body motion (i.e.\ stress-free) due to the surrounding pore space 
	\item There is no observance of a macroscopic bending stiffness
	\item The reloading response was found to exhibit a strong dependency on $\stretchMacroMin$, which governs the degree of pore space closure during unloading, which increases intermolecular interactions
	
	\item Hysteretic behaviour of a craze under cyclic loading is generally also observed in experimental studies \cite{Kambour.1969, Hoare.1972}
\end{itemize}
}

\Notes{
Key findings from bulk-craze interaction
\begin{itemize}
	\item For different craze-bulk compositions, the stress-strain can be collapsed if normalised with the maximum deformation during loading, i.e.\ the craze governs the macroscopic response independent of its here investigated length
	\item This is attributed to a much lower craze stiffness with respect to $\stretchMacro$ compared to the bulk stiffness
	\item A consequence is the derivation of a simple continuum-micromechanical model to estimate the current craze fibril length, which can be used in constitutive models
\end{itemize}
}

\Notes{
Two valuable contributions for future work include
\begin{itemize}
	\item a constitutive craze fibril model which allows to incorporate the response on a continuum scale 
	\item and strain controlled experimental investigations of craze fibrils, since stress-controlled experiments \cite{Kambour.1969} are not able to capture mechanical insight into pore space closure and the subsequent reloading response.
\end{itemize}
}

%% file: sections/90_acknowledgement.tex
\begin{acknowledgement}
	We thank Siteng Zhang for providing the equilibrated melt configuration. 
	Partial work was performed throughout a research stay of TL at the University of British Columbia, which was financially supported by the Karlsruhe House of Young Scientists (KHYS). 
	Computational support from the state of Baden-Württemberg through bwHPC is gratefully acknowledged. 
	
%	Please use ``The authors thank \ldots'' rather than ``The
%	authors would like to thank \ldots''.
%	
%	The author thanks Mats Dahlgren for version one of \textsf{achemso},
%	and Donald Arseneau for the code taken from \textsf{cite} to move
%	citations after punctuation. Many users have provided feedback on the
%	class, which is reflected in all of the different demonstrations
%	shown in this document.
	
\end{acknowledgement}

%% file: sections/91_suppinfo.tex
\begin{suppinfo} \label{sec_suppinfo}
The file (PDF) contains details on the animations and additional data presenting
\begin{itemize}
	\item influence of deformation rate and temperature,
	\item stress evolution for multiple loading cycles,
	\item stress decomposition for $\stretchMacroMin=1$ and $\stretchMacroMin=2$ and
	\item evolution of craze length components for $\stretchMacroMin=1$ and $\stretchMacroMin=2$ in a non-normalised manner
\end{itemize}

MS1: Animation visualizing the cyclic craze response for $\stretchMacroMin=1$ (MP4)

MS2: Animation visualizing the cyclic craze response for $\stretchMacroMin=2$ (MP4)

MS3: Animation visualizing the cyclic craze response for $\stretchMacroMin=5$ (MP4)

\end{suppinfo}

%% file: main.bbl
\providecommand{\latin}[1]{#1}
\makeatletter
\providecommand{\doi}
  {\begingroup\let\do\@makeother\dospecials
  \catcode`\{=1 \catcode`\}=2 \doi@aux}
\providecommand{\doi@aux}[1]{\endgroup\texttt{#1}}
\makeatother
\providecommand*\mcitethebibliography{\thebibliography}
\csname @ifundefined\endcsname{endmcitethebibliography}
  {\let\endmcitethebibliography\endthebibliography}{}
\begin{mcitethebibliography}{41}
\providecommand*\natexlab[1]{#1}
\providecommand*\mciteSetBstSublistMode[1]{}
\providecommand*\mciteSetBstMaxWidthForm[2]{}
\providecommand*\mciteBstWouldAddEndPuncttrue
  {\def\EndOfBibitem{\unskip.}}
\providecommand*\mciteBstWouldAddEndPunctfalse
  {\let\EndOfBibitem\relax}
\providecommand*\mciteSetBstMidEndSepPunct[3]{}
\providecommand*\mciteSetBstSublistLabelBeginEnd[3]{}
\providecommand*\EndOfBibitem{}
\mciteSetBstSublistMode{f}
\mciteSetBstMaxWidthForm{subitem}{(\alph{mcitesubitemcount})}
\mciteSetBstSublistLabelBeginEnd
  {\mcitemaxwidthsubitemform\space}
  {\relax}
  {\relax}

\bibitem[Kramer and Berger(1990)Kramer, and Berger]{Kramer.1990}
Kramer,~E.~J.; Berger,~L.~L. In \emph{Crazing in Polymers}; Kausch,~H.~H., Ed.;
  Advances in Polymer Science; Springer and {Central Book Services New Zealand
  [distributor]}: Berlin and Mitcham, VIC, Australia, 1990; Vol. 91/92; pp
  1--68\relax
\mciteBstWouldAddEndPuncttrue
\mciteSetBstMidEndSepPunct{\mcitedefaultmidpunct}
{\mcitedefaultendpunct}{\mcitedefaultseppunct}\relax
\EndOfBibitem
\bibitem[Kambour(1973)]{Kambour.1973}
Kambour,~R.~P. A review of crazing and fracture in thermoplastics.
  \emph{Journal of Polymer Science: Macromolecular Reviews} \textbf{1973},
  \emph{7}, 1--154\relax
\mciteBstWouldAddEndPuncttrue
\mciteSetBstMidEndSepPunct{\mcitedefaultmidpunct}
{\mcitedefaultendpunct}{\mcitedefaultseppunct}\relax
\EndOfBibitem
\bibitem[Kausch(1983)]{Kausch.1983}
Kausch,~H.-H., Ed. \emph{Crazing in polymers}; Advances in Polymer Science;
  Springer: Berlin, 1983; Vol. 52/53\relax
\mciteBstWouldAddEndPuncttrue
\mciteSetBstMidEndSepPunct{\mcitedefaultmidpunct}
{\mcitedefaultendpunct}{\mcitedefaultseppunct}\relax
\EndOfBibitem
\bibitem[Kausch(1990)]{Kausch.1990}
Kausch,~H.~H., Ed. \emph{Crazing in Polymers}; Advances in Polymer Science;
  Springer and {Central Book Services New Zealand [distributor]}: Berlin and
  Mitcham, VIC, Australia, 1990\relax
\mciteBstWouldAddEndPuncttrue
\mciteSetBstMidEndSepPunct{\mcitedefaultmidpunct}
{\mcitedefaultendpunct}{\mcitedefaultseppunct}\relax
\EndOfBibitem
\bibitem[Haward and Young(1997)Haward, and Young]{Haward.1997}
Haward,~R.~N.; Young,~R.~J. \emph{The Physics of Glassy Polymers}; {Springer
  Netherlands}: Dordrecht, 1997\relax
\mciteBstWouldAddEndPuncttrue
\mciteSetBstMidEndSepPunct{\mcitedefaultmidpunct}
{\mcitedefaultendpunct}{\mcitedefaultseppunct}\relax
\EndOfBibitem
\bibitem[Tijssens \latin{et~al.}(2000)Tijssens, {van der Giessen}, and
  Sluys]{Tijssens.2000a}
Tijssens,~M.; {van der Giessen},~E.; Sluys,~L.~J. Modeling of crazing using a
  cohesive surface methodology. \emph{Mechanics of Materials} \textbf{2000},
  \emph{32}, 19--35\relax
\mciteBstWouldAddEndPuncttrue
\mciteSetBstMidEndSepPunct{\mcitedefaultmidpunct}
{\mcitedefaultendpunct}{\mcitedefaultseppunct}\relax
\EndOfBibitem
\bibitem[Estevez \latin{et~al.}(2000)Estevez, Tijssens, and {van der
  Giessen}]{Estevez.2000}
Estevez,~R.; Tijssens,~M.; {van der Giessen},~E. Modeling of the competition
  between shear yielding and crazing in glassy polymers. \emph{Journal of the
  Mechanics and Physics of Solids} \textbf{2000}, \emph{48}, 2585--2617\relax
\mciteBstWouldAddEndPuncttrue
\mciteSetBstMidEndSepPunct{\mcitedefaultmidpunct}
{\mcitedefaultendpunct}{\mcitedefaultseppunct}\relax
\EndOfBibitem
\bibitem[Socrate \latin{et~al.}(2001)Socrate, Boyce, and Lazzeri]{Socrate.2001}
Socrate,~S.; Boyce,~M.~C.; Lazzeri,~A. A micromechanical model for multiple
  crazing in high impact polystyrene. \emph{Mechanics of Materials}
  \textbf{2001}, \emph{33}, 155--175\relax
\mciteBstWouldAddEndPuncttrue
\mciteSetBstMidEndSepPunct{\mcitedefaultmidpunct}
{\mcitedefaultendpunct}{\mcitedefaultseppunct}\relax
\EndOfBibitem
\bibitem[Gearing and Anand(2004)Gearing, and Anand]{Gearing.2004}
Gearing,~B.~P.; Anand,~L. On modeling the deformation and fracture response of
  glassy polymers due to shear-yielding and crazing. \emph{International
  Journal of Solids and Structures} \textbf{2004}, \emph{41}, 3125--3150\relax
\mciteBstWouldAddEndPuncttrue
\mciteSetBstMidEndSepPunct{\mcitedefaultmidpunct}
{\mcitedefaultendpunct}{\mcitedefaultseppunct}\relax
\EndOfBibitem
\bibitem[Basu \latin{et~al.}(2005)Basu, Mahajan, and {van der
  Giessen}]{Basu.2005}
Basu,~S.; Mahajan,~D.~K.; {van der Giessen},~E. Micromechanics of the growth of
  a craze fibril in glassy polymers. \emph{Polymer} \textbf{2005}, \emph{46},
  7504--7518\relax
\mciteBstWouldAddEndPuncttrue
\mciteSetBstMidEndSepPunct{\mcitedefaultmidpunct}
{\mcitedefaultendpunct}{\mcitedefaultseppunct}\relax
\EndOfBibitem
\bibitem[Seelig(2008)]{Seelig.2008}
Seelig,~T. Computational modeling of deformation mechanisms and failure in
  thermoplastic multilayer composites. \emph{Composites Science and Technology}
  \textbf{2008}, \emph{68}, 1198--1208\relax
\mciteBstWouldAddEndPuncttrue
\mciteSetBstMidEndSepPunct{\mcitedefaultmidpunct}
{\mcitedefaultendpunct}{\mcitedefaultseppunct}\relax
\EndOfBibitem
\bibitem[Helbig \latin{et~al.}(2016)Helbig, {van der Giessen}, Clausen, and
  Seelig]{Helbig.2016}
Helbig,~M.; {van der Giessen},~E.; Clausen,~A.~H.; Seelig,~T.
  Continuum-micromechanical modeling of distributed crazing in rubber-toughened
  polymers. \emph{European Journal of Mechanics - A/Solids} \textbf{2016},
  \emph{57}, 108--120\relax
\mciteBstWouldAddEndPuncttrue
\mciteSetBstMidEndSepPunct{\mcitedefaultmidpunct}
{\mcitedefaultendpunct}{\mcitedefaultseppunct}\relax
\EndOfBibitem
\bibitem[Baljon and Robbins(2001)Baljon, and Robbins]{Baljon.2001}
Baljon,~A. R.~C.; Robbins,~M.~O. Simulations of Crazing in Polymer Glasses:
  Effect of Chain Length and Surface Tension. \emph{Macromolecules}
  \textbf{2001}, \emph{34}, 4200--4209\relax
\mciteBstWouldAddEndPuncttrue
\mciteSetBstMidEndSepPunct{\mcitedefaultmidpunct}
{\mcitedefaultendpunct}{\mcitedefaultseppunct}\relax
\EndOfBibitem
\bibitem[Rottler and Robbins(2002)Rottler, and Robbins]{Rottler.2002.jamming}
Rottler,~J.; Robbins,~M.~O. Jamming under Tension in Polymer Crazes.
  \emph{Phys. Rev. Lett.} \textbf{2002}, \emph{89}, 195501\relax
\mciteBstWouldAddEndPuncttrue
\mciteSetBstMidEndSepPunct{\mcitedefaultmidpunct}
{\mcitedefaultendpunct}{\mcitedefaultseppunct}\relax
\EndOfBibitem
\bibitem[Rottler \latin{et~al.}(2002)Rottler, Barsky, and
  Robbins]{Rottler.2002}
Rottler,~J.; Barsky,~S.; Robbins,~M.~O. Cracks and Crazes: On Calculating the
  Macroscopic Fracture Energy of Glassy Polymers from Molecular Simulations.
  \emph{Phys. Rev. Lett.} \textbf{2002}, \emph{89}, 148304\relax
\mciteBstWouldAddEndPuncttrue
\mciteSetBstMidEndSepPunct{\mcitedefaultmidpunct}
{\mcitedefaultendpunct}{\mcitedefaultseppunct}\relax
\EndOfBibitem
\bibitem[Rottler and Robbins(2003)Rottler, and Robbins]{Rottler.2003}
Rottler,~J.; Robbins,~M.~O. Growth, microstructure, and failure of crazes in
  glassy polymers. \emph{Physical review. E, Statistical, nonlinear, and soft
  matter physics} \textbf{2003}, \emph{68}, 011801\relax
\mciteBstWouldAddEndPuncttrue
\mciteSetBstMidEndSepPunct{\mcitedefaultmidpunct}
{\mcitedefaultendpunct}{\mcitedefaultseppunct}\relax
\EndOfBibitem
\bibitem[Mahajan \latin{et~al.}(2010)Mahajan, Singh, and Basu]{Basu.2010}
Mahajan,~D.~K.; Singh,~B.; Basu,~S. Void nucleation and disentanglement in
  glassy amorphous polymers. \emph{Phys. Rev. E} \textbf{2010}, \emph{82},
  011803\relax
\mciteBstWouldAddEndPuncttrue
\mciteSetBstMidEndSepPunct{\mcitedefaultmidpunct}
{\mcitedefaultendpunct}{\mcitedefaultseppunct}\relax
\EndOfBibitem
\bibitem[Toepperwein and de~Pablo(2011)Toepperwein, and
  de~Pablo]{toepperwein_2011}
Toepperwein,~G.~N.; de~Pablo,~J.~J. Cavitation and Crazing in Rod-Containing
  Nanocomposites. \emph{Macromolecules} \textbf{2011}, \emph{44},
  5498--5509\relax
\mciteBstWouldAddEndPuncttrue
\mciteSetBstMidEndSepPunct{\mcitedefaultmidpunct}
{\mcitedefaultendpunct}{\mcitedefaultseppunct}\relax
\EndOfBibitem
\bibitem[Mahajan and Hartmaier(2012)Mahajan, and Hartmaier]{mahajan_2012}
Mahajan,~D.~K.; Hartmaier,~A. Mechanisms of crazing in glassy polymers revealed
  by molecular dynamics simulations. \emph{Phys. Rev. E} \textbf{2012},
  \emph{86}, 021802\relax
\mciteBstWouldAddEndPuncttrue
\mciteSetBstMidEndSepPunct{\mcitedefaultmidpunct}
{\mcitedefaultendpunct}{\mcitedefaultseppunct}\relax
\EndOfBibitem
\bibitem[Venkatesan and Basu(2015)Venkatesan, and Basu]{Venkatesan.2015}
Venkatesan,~S.; Basu,~S. Investigations into crazing in glassy amorphous
  polymers through molecular dynamics simulations. \emph{Journal of the
  Mechanics and Physics of Solids} \textbf{2015}, \emph{77}, 123--145\relax
\mciteBstWouldAddEndPuncttrue
\mciteSetBstMidEndSepPunct{\mcitedefaultmidpunct}
{\mcitedefaultendpunct}{\mcitedefaultseppunct}\relax
\EndOfBibitem
\bibitem[Ge \latin{et~al.}(2014)Ge, Grest, and Robbins]{Ge.2014}
Ge,~T.; Grest,~G.~S.; Robbins,~M.~O. Tensile Fracture of Welded Polymer
  Interfaces: Miscibility, Entanglements, and Crazing. \emph{Macromolecules}
  \textbf{2014}, \emph{47}, 6982--6989\relax
\mciteBstWouldAddEndPuncttrue
\mciteSetBstMidEndSepPunct{\mcitedefaultmidpunct}
{\mcitedefaultendpunct}{\mcitedefaultseppunct}\relax
\EndOfBibitem
\bibitem[Ge \latin{et~al.}(2017)Ge, Tzoumanekas, Anogiannakis, Hoy, and
  Robbins]{Ge.2017}
Ge,~T.; Tzoumanekas,~C.; Anogiannakis,~S.~D.; Hoy,~R.~S.; Robbins,~M.~O.
  Entanglements in Glassy Polymer Crazing: Cross-Links or Tubes?
  \emph{Macromolecules} \textbf{2017}, \emph{50}, 459--471\relax
\mciteBstWouldAddEndPuncttrue
\mciteSetBstMidEndSepPunct{\mcitedefaultmidpunct}
{\mcitedefaultendpunct}{\mcitedefaultseppunct}\relax
\EndOfBibitem
\bibitem[Wang \latin{et~al.}(2022)Wang, in~{'t Veld}, Robbins, and
  Ge]{Wang.2022}
Wang,~J.; in~{'t Veld},~P.~J.; Robbins,~M.~O.; Ge,~T. Effects of
  Coarse-Graining on Molecular Simulation of Craze Formation in Polymer Glass.
  \emph{Macromolecules} \textbf{2022}, \emph{55}, 1267--1278\relax
\mciteBstWouldAddEndPuncttrue
\mciteSetBstMidEndSepPunct{\mcitedefaultmidpunct}
{\mcitedefaultendpunct}{\mcitedefaultseppunct}\relax
\EndOfBibitem
\bibitem[Dietz \latin{et~al.}(2022)Dietz, Nan, and Hoy]{dietz_2022}
Dietz,~J.~D.; Nan,~K.; Hoy,~R.~S. Unexpected Ductility in Semiflexible Polymer
  Glasses with Entanglement Length Equal to Their Kuhn Length. \emph{Phys. Rev.
  Lett.} \textbf{2022}, \emph{129}, 127801\relax
\mciteBstWouldAddEndPuncttrue
\mciteSetBstMidEndSepPunct{\mcitedefaultmidpunct}
{\mcitedefaultendpunct}{\mcitedefaultseppunct}\relax
\EndOfBibitem
\bibitem[Nan and Hoy(2023)Nan, and Hoy]{nan_2023}
Nan,~K.; Hoy,~R.~S. Craze Extension Ratio of Semiflexible Polymer Glasses.
  \emph{Macromolecules} \textbf{2023}, \emph{56}, 8369--8375\relax
\mciteBstWouldAddEndPuncttrue
\mciteSetBstMidEndSepPunct{\mcitedefaultmidpunct}
{\mcitedefaultendpunct}{\mcitedefaultseppunct}\relax
\EndOfBibitem
\bibitem[D{\"o}ll and K{\"o}ncz{\"o}l(1990)D{\"o}ll, and
  K{\"o}ncz{\"o}l]{Doell.1990}
D{\"o}ll,~W.; K{\"o}ncz{\"o}l,~L. In \emph{Crazing in Polymers}; Kausch,~H.~H.,
  Ed.; Advances in Polymer Science; Springer and {Central Book Services New
  Zealand [distributor]}: Berlin and Mitcham, VIC, Australia, 1990; Vol. 91/92;
  pp 137--214\relax
\mciteBstWouldAddEndPuncttrue
\mciteSetBstMidEndSepPunct{\mcitedefaultmidpunct}
{\mcitedefaultendpunct}{\mcitedefaultseppunct}\relax
\EndOfBibitem
\bibitem[Schirrer(1990)]{Schirrer.1990}
Schirrer,~R. In \emph{Crazing in Polymers Vol. 2}; Kausch,~H.~H., Ed.; Springer
  Berlin Heidelberg: Berlin, Heidelberg, 1990; pp 215--261\relax
\mciteBstWouldAddEndPuncttrue
\mciteSetBstMidEndSepPunct{\mcitedefaultmidpunct}
{\mcitedefaultendpunct}{\mcitedefaultseppunct}\relax
\EndOfBibitem
\bibitem[Takemori(1990)]{Takemori.1990}
Takemori,~M.~T. In \emph{Crazing in Polymers}; Kausch,~H.~H., Ed.; Advances in
  Polymer Science; Springer and {Central Book Services New Zealand
  [distributor]}: Berlin and Mitcham, VIC, Australia, 1990; Vol. 91/92; pp
  263--300\relax
\mciteBstWouldAddEndPuncttrue
\mciteSetBstMidEndSepPunct{\mcitedefaultmidpunct}
{\mcitedefaultendpunct}{\mcitedefaultseppunct}\relax
\EndOfBibitem
\bibitem[Skibo \latin{et~al.}(1977)Skibo, Hertzberg, Manson, and
  Kim]{Skibo.1977}
Skibo,~M.~D.; Hertzberg,~R.~W.; Manson,~J.~A.; Kim,~S.~L. On the generality of
  discontinuous fatigue crack growth in glassy polymers. \emph{Journal of
  Materials Science} \textbf{1977}, \emph{12}, 531--542\relax
\mciteBstWouldAddEndPuncttrue
\mciteSetBstMidEndSepPunct{\mcitedefaultmidpunct}
{\mcitedefaultendpunct}{\mcitedefaultseppunct}\relax
\EndOfBibitem
\bibitem[K{\"o}ncz{\"o}l \latin{et~al.}(1990)K{\"o}ncz{\"o}l, D{\"o}ll, and
  Bevan]{Koenczoel.1990}
K{\"o}ncz{\"o}l,~L.; D{\"o}ll,~W.; Bevan,~L. Mechanisms and micromechanics of
  fatigue crack propagation in glassy thermoplastics. \emph{Colloid {\&}
  Polymer Science} \textbf{1990}, \emph{268}, 814--822\relax
\mciteBstWouldAddEndPuncttrue
\mciteSetBstMidEndSepPunct{\mcitedefaultmidpunct}
{\mcitedefaultendpunct}{\mcitedefaultseppunct}\relax
\EndOfBibitem
\bibitem[Laschuetza and Seelig(2024)Laschuetza, and Seelig]{Laschuetza.2024}
Laschuetza,~T.; Seelig,~T. A continuum-micromechanical model for crazing in
  glassy polymers under cyclic loading. \emph{Mechanics of Materials}
  \textbf{2024}, \emph{189}, 104901\relax
\mciteBstWouldAddEndPuncttrue
\mciteSetBstMidEndSepPunct{\mcitedefaultmidpunct}
{\mcitedefaultendpunct}{\mcitedefaultseppunct}\relax
\EndOfBibitem
\bibitem[Kambour and Kopp(1969)Kambour, and Kopp]{Kambour.1969}
Kambour,~R.~P.; Kopp,~R.~W. Cyclic stress--strain behavior of the dry
  polycarbonate craze. \emph{Journal of Polymer Science Part A-2: Polymer
  Physics} \textbf{1969}, \emph{7}, 183--200\relax
\mciteBstWouldAddEndPuncttrue
\mciteSetBstMidEndSepPunct{\mcitedefaultmidpunct}
{\mcitedefaultendpunct}{\mcitedefaultseppunct}\relax
\EndOfBibitem
\bibitem[Hoare and Hull(1972)Hoare, and Hull]{Hoare.1972}
Hoare,~J.; Hull,~D. Craze yielding and stress-strain characteristics of crazes
  in polystyrene. \emph{The Philosophical Magazine: A Journal of Theoretical
  Experimental and Applied Physics} \textbf{1972}, \emph{26}, 443--455\relax
\mciteBstWouldAddEndPuncttrue
\mciteSetBstMidEndSepPunct{\mcitedefaultmidpunct}
{\mcitedefaultendpunct}{\mcitedefaultseppunct}\relax
\EndOfBibitem
\bibitem[Kremer and Grest(1990)Kremer, and Grest]{Kremer.1990}
Kremer,~K.; Grest,~G.~S. Dynamics of entangled linear polymer melts: A
  molecular‐dynamics simulation. \emph{The Journal of Chemical Physics}
  \textbf{1990}, \emph{92}, 5057--5086\relax
\mciteBstWouldAddEndPuncttrue
\mciteSetBstMidEndSepPunct{\mcitedefaultmidpunct}
{\mcitedefaultendpunct}{\mcitedefaultseppunct}\relax
\EndOfBibitem
\bibitem[Plimpton(1995)]{LAMMPS.1995}
Plimpton,~S. Fast Parallel Algorithms for Short-Range Molecular Dynamics.
  \emph{Journal of Computational Physics} \textbf{1995}, \emph{117},
  1--19\relax
\mciteBstWouldAddEndPuncttrue
\mciteSetBstMidEndSepPunct{\mcitedefaultmidpunct}
{\mcitedefaultendpunct}{\mcitedefaultseppunct}\relax
\EndOfBibitem
\bibitem[Thompson \latin{et~al.}(2022)Thompson, Aktulga, Berger, Bolintineanu,
  Brown, Crozier, in~'t Veld, Kohlmeyer, Moore, Nguyen, Shan, Stevens,
  Tranchida, Trott, and Plimpton]{LAMMPS.2022}
Thompson,~A.~P.; Aktulga,~H.~M.; Berger,~R.; Bolintineanu,~D.~S.; Brown,~W.~M.;
  Crozier,~P.~S.; in~'t Veld,~P.~J.; Kohlmeyer,~A.; Moore,~S.~G.;
  Nguyen,~T.~D.; Shan,~R.; Stevens,~M.~J.; Tranchida,~J.; Trott,~C.;
  Plimpton,~S.~J. {LAMMPS} - a flexible simulation tool for particle-based
  materials modeling at the atomic, meso, and continuum scales. \emph{Comp.
  Phys. Comm.} \textbf{2022}, \emph{271}, 108171\relax
\mciteBstWouldAddEndPuncttrue
\mciteSetBstMidEndSepPunct{\mcitedefaultmidpunct}
{\mcitedefaultendpunct}{\mcitedefaultseppunct}\relax
\EndOfBibitem
\bibitem[Auhl \latin{et~al.}(2003)Auhl, Everaers, Grest, Kremer, and
  Plimpton]{Auhl.2003}
Auhl,~R.; Everaers,~R.; Grest,~G.~S.; Kremer,~K.; Plimpton,~S.~J.
  {Equilibration of long chain polymer melts in computer simulations}.
  \emph{The Journal of Chemical Physics} \textbf{2003}, \emph{119},
  12718--12728\relax
\mciteBstWouldAddEndPuncttrue
\mciteSetBstMidEndSepPunct{\mcitedefaultmidpunct}
{\mcitedefaultendpunct}{\mcitedefaultseppunct}\relax
\EndOfBibitem
\bibitem[Hoy and Robbins(2005)Hoy, and Robbins]{Hoy.2005}
Hoy,~R.~S.; Robbins,~M.~O. Effect of equilibration on primitive path analyses
  of entangled polymers. \emph{Phys. Rev. E} \textbf{2005}, \emph{72},
  061802\relax
\mciteBstWouldAddEndPuncttrue
\mciteSetBstMidEndSepPunct{\mcitedefaultmidpunct}
{\mcitedefaultendpunct}{\mcitedefaultseppunct}\relax
\EndOfBibitem
\bibitem[Stukowski(2009)]{OVITO.2010}
Stukowski,~A. Visualization and analysis of atomistic simulation data with
  OVITO–the Open Visualization Tool. \emph{Modelling and Simulation in
  Materials Science and Engineering} \textbf{2009}, \emph{18}, 015012\relax
\mciteBstWouldAddEndPuncttrue
\mciteSetBstMidEndSepPunct{\mcitedefaultmidpunct}
{\mcitedefaultendpunct}{\mcitedefaultseppunct}\relax
\EndOfBibitem
\bibitem[Bevan \latin{et~al.}(1986)Bevan, Doell, and Koenczoel]{Bevan.1986}
Bevan,~L.; Doell,~W.; Koenczoel,~L. Micromechanics of a craze zone at the tip
  of a stationary crack. \emph{Journal of Polymer Science Part B: Polymer
  Physics} \textbf{1986}, \emph{24}, 2433--2444\relax
\mciteBstWouldAddEndPuncttrue
\mciteSetBstMidEndSepPunct{\mcitedefaultmidpunct}
{\mcitedefaultendpunct}{\mcitedefaultseppunct}\relax
\EndOfBibitem
\end{mcitethebibliography}
